\documentclass[useAMS,usenatbib]{mn2e}
\usepackage{graphicx}
\usepackage{natbib}
\usepackage{lscape}
\usepackage{longtable}
\usepackage{subfig}
\usepackage{amssymb,amsmath}
\usepackage[T1]{fontenc}
\usepackage{hyperref}
\usepackage{amsmath}
\usepackage{breakurl}
\usepackage{times}
\usepackage{morefloats}
\usepackage{bm}
\usepackage{flafter}

\newcommand{\wise}{\textit{WISE}}

\newcommand{\planck}{\textit{Planck}}

\newcommand\ion[2]{#1$\;${\small\rmfamily\@Roman{#2}}\relax}%

\def\lsim{\lower0.3em\hbox{$\,\buildrel <\over\sim\,$}}
\def\gsim{\lower0.3em\hbox{$\,\buildrel >\over\sim\,$}}

\title[Halo masses of \wise\ quasars]{The characteristic halo masses of half-a-million \emph{WISE}-selected quasars}
\author[DiPompeo et al.]{M.A. DiPompeo$^{1}$, R.C. Hickox$^1$, S. Eftekharzadeh$^2$, A.D. Myers$^2$ \\
$^1$ Department of Physics and Astronomy, Dartmouth College, 6127 Wilder Laboratory, Hanover, NH 03755, USA  \\
$^2$ Department of Physics and Astronomy 3905, University of Wyoming, 1000 E. University, Laramie, WY 82071, USA}

\begin{document}
\date{Accepted 2017 May 15; Received 2017 May 11; in original form 2017 March 24}

\pagerange{\pageref{firstpage}--\pageref{lastpage}} \pubyear{2015}

\maketitle

\label{firstpage}

\begin{abstract}
Recent work has found evidence for a difference in the bias and dark matter halo masses of \wise-selected obscured and unobscured quasars, implying a distinction between these populations beyond random line-of-sight effects.  However, the significance of this difference in the most up-to-date measurements is relatively weak, at $\sim$2$\sigma$ for individual measurements but bolstered by agreement from different techniques, including angular clustering and cross-correlations with cosmic microwave background (CMB) lensing maps.  Here, we expand the footprint of previous work, aiming to improve the precision of both methods. In this larger area we correct for position dependent selection effects, in particular fluctuations of the \wise-selected quasar density as a function of Galactic latitude.  We also measure the cross-correlation of the obscured and unobscured samples and confirm that they are well-matched in redshift, both centred at $z=1$. Combined with very similar detection fractions and magnitude distributions in the long-wavelength \wise\ bands, this redshift match strongly supports the fact that IR selection identifies obscured and unobscured quasars of similar bolometric luminosity. Finally, we perform cross-correlations with confirmed spectroscopic quasars, again confirming the results from other methods --- obscured quasars reside in haloes a factor of 3 times more massive than unobscured quasars.  This difference is significant at the $\sim$5$\sigma$ level when the measurements are combined, strong support for the idea that obscuration in at least some quasars is tied to the larger environment, and may have an evolutionary component.
\end{abstract}


\section{INTRODUCTION} \label{sec:intro}
With the development of simple infrared (IR) quasar selection criteria \citep[e.g.][]{2004ApJS..154..166L, Stern:2005p2563, 2012ApJ...753...30S, 2012MNRAS.426.3271M, 2013ApJ...772...26A}, large samples can now be assembled across the entire sky from the \emph{Wide-field Infrared Survey Explorer} \citep[\emph{WISE};][]{2010AJ....140.1868W}. Critically, selecting quasars in the mid-IR identifies obscured quasars, which have been previously poorly represented in statistical samples, while still also selecting the much better studied unobscured population.  While the bolometric luminosities of the two classes selected this way are quite similar \citep{2007ApJ...671.1365H}, large obscuring columns of dust make obscured quasars difficult to identify photometrically in optical surveys, such as the Sloan Digital Sky Survey \citep[SDSS;][]{2000AJ....120.1579Y}, despite comprising roughly half of the quasar population \citep{2013ApJ...772...26A}.  The ability of mid-IR radiation to penetrate this dust provides a unique tool for exploring these ``hidden monsters.''

With large populations of both obscured and unobscured quasars selected in a similar manner, we can now deeply explore questions regarding the nature of obscured quasars in the context of both AGN unification models \citep[e.g.][]{1993ARA&A..31..473A} and evolutionary models of quasars and galaxies \citep[e.g.][]{2008ApJS..175..356H, 2009MNRAS.394.1109C}.  In particular, the location and distribution of the obscuring material in the most luminous quasars, such as those selected with simple colour-cuts from relatively shallow \emph{WISE} data, is still a matter of debate.  The simplest unification models ascribe all of the obscuration to random orientations and an axis-symmetric structure just beyond the accretion disk commonly known as the ``dusty torus'', the properties of which may depend on other factors such as luminosity \citep[e.g.][]{1991MNRAS.252..586L, 2015ARA&A..53..365N}.  This explanation implies that the bulk properties of the unobscured and obscured populations on larger scales should be the same, including those of their host galaxies, dark matter haloes, and environment more generally.

However, obscuration can also occur due to dust in the host galaxy \citep[e.g.][]{2004ApJ...611L..85P, 2012ApJ...755....5G, 2017MNRAS.465.4348B}.  While this can also be due to the randomness of our line of sight, models that invoke galaxy mergers as a significant fueling mechanism for black hole growth suggest that large-scale, high covering fraction obscuration can occur naturally in an evolutionary quasar sequence \citep[e.g.][]{1988ApJ...325...74S, 2008ApJS..175..356H}. In such a scenario, where obscured quasars are a phase in the overall quasar life cycle, differences in the host galaxies \citep[e.g.][]{2015ApJ...802...50C} and larger scale environments may be present \citep[e.g.][]{2015MNRAS.451L..35E}

Over the last several years, much work has been done to test these models by measuring the characteristic dark matter halo masses of IR-selected obscured and unobscured quasars \citep[see ][hereafter D16, for a detailed review of the IR-selection of quasars and the use of such samples to probe large-scale structure]{2016MNRAS.456..924D}. While the results have varied depending on the methodology and precise sample selection, most have agreed that IR-selected obscured quasars reside in haloes at least a factor of a few times more massive than their unobscured counterparts \citep[][]{2011ApJ...731..117H, 2014ApJ...789...44D, 2014MNRAS.442.3443D, 2015MNRAS.446.3492D, 2016MNRAS.456..924D}.  Using simple relationships between halo mass, galaxy mass, and black hole mass, \citet{2017MNRAS.464.3526D} showed that a simple evolutionary trend from obscured to unobscured quasar, with black hole growth lagging behind that of the rest of the system, could broadly reproduce the measured halo mass difference, as well as explain some discrepancies in the literature \citep[i.e.\ the similar IR-selected obscured/unobscured quasar halo masses at lower redshift and luminosity;][]{2016ApJ...821...55M}.

Though the halo mass difference was still present in the most recent measurements of D16, the significance was reduced to $\sim$2$\sigma$ after employing the latest photometric samples and CMB lensing maps.  While this difference is likely a lower limit on the potentially evolutionarily distinct obscured population \citep[due to a subset being unified by orientation;][]{2016MNRAS.460..175D}, the measurements thus far have relied on a portion of the total available area, where systematics in the survey data are less prevalent.  Given the important implications of any potential halo mass difference between the populations and the current state of the measurements, it is prudent to expand the footprint, increasing the sample sizes to their full potential for the most precise possible measurement.  That, along with exploring cross-correlations, is our goal in this work.

All models and measured properties use a cosmology of $H_0 = 70.2$ km s$^{-1}$ Mpc$^{-1}$, $\Omega_{\textrm{M}} = \Omega_{\textrm{CDM}} + \Omega_{\textrm{b}} = 0.229 + 0.046 = 0.275$, $\Omega_{\Lambda} = 0.725$, and $\sigma_8 = 0.82$ \citep{2011ApJS..192...18K}.  Magnitudes are given in the Vega system unless otherwise specified.

\section{Quasar Sample} \label{sec:data}
\subsection{\emph{WISE} Quasar Selection}
We make our initial sample selection using both the original ALLSKY and the updated ALLWISE catalogs from \emph{WISE}.  The \emph{WISE} mission mapped the entire sky in four bands: 3.4, 4.6, 12, and 22 $\mu$m, designated as $W1$, $W2$, $W3$, and $W4$, respectively.  The two shortest wavelength bands are the most sensitive and highest resolution ($\sim$0.1 mJy 5$\sigma$ sensitivity compared to $\sim$1 mJy, and $\sim$6 arcsec versus $\sim$10 arcsec resolution), and are the most efficient for selecting luminous quasars \citep{2012ApJ...753...30S}.  The ALLSKY catalog was the first full source catalogue release in March 2012, and was supplanted by the ALLWISE release in November 2013, which included new imaging in the $W1$ and $W2$ bands and an improved data calibration pipeline.  

D16 provided a detailed analysis of quasar samples selected from these two catalogs in the context of the measurements performed here (angular clustering in Section~\ref{sec:clustering} and CMB lensing cross-correlations in Section~\ref{sec:lensing}).  While they found that the ALLWISE survey provides a slightly brighter and bluer but less-contaminated sample, there were potentially problematic systematic correlations in the source distribution with Moon contamination that were not present in the ALLSKY catalog.  In order to provide the benefits of both and a conservatively generated sample, we adopt their approach and select objects that satisfy the \citet{2012ApJ...753...30S} selection criteria of $W1 - W2 > 0.8$ and $W2 < 15.05$ in \emph{both} catalogs.

After selecting objects from the catalog data, we apply the same ``mask'' to the data as D16, which discards regions of likely contamination by defining holes in the survey footprint.  We refer the reader to D16 for complete details, but in general we remove the Galactic plane within $|b| \le 20$ degrees, regions of high Galactic extinction where $A_g > 0.18$, regions of high Moon contamination, regions around clustered data flagged as spurious in the \emph{WISE} catalogs \citep[using \textsc{cc\_flags}; see][]{2014MNRAS.442.3443D}, regions of poor \emph{WISE} photometric quality (based on the \textsc{ph\_qual} flag), and areas around bright stars as catalogued by SDSS.  These masks are generated for both the ALLSKY and ALLWISE catalogs and both are applied --- we provide \textsc{mangle} \citep{2008MNRAS.387.1391S} polygon files of the mask at \url{http://faraday.uwyo.edu/~admyers/wisemask2017/wisemask.html}. 

After applying the mask, we are left with 581,728 \emph{WISE}-selected quasars over 12,745 deg$^2$.  The distribution of this sample on the sky is shown in red in Figure~\ref{fig:samples}.

\subsection{Obscured and Unobscured Quasars}
While the IR data alone can select quasars, additional optical information is needed to divide the sample into obscured and unobscured populations.  Here we follow \citet{2007ApJ...671.1365H} and exploit the bimodal distribution of quasar optical-IR colours, using $r-W2=6$ to divide the samples\footnote{We stress that this definition is different from the more traditional type 1 (unobscured) and type 2 (obscured) quasar definitions, which are based on the presence or absence of broad emission lines in spectra.} (see also Hickox et al. 2017, in prep).  

We gather optical $r$-band magnitudes from data release 8 of the SDSS \citep[DR8\footnote{Note that the astrometric error in SDSS DR8 that was subsequently corrected in DR9 \citep{2012ApJS..203...21A} does not impact our matching, given the comparatively low resolution of the \wise\ imaging. All of our objects are matched to the same optical counterpart in DR8 and DR9.}, adopting \textsc{psf\_mags} though the use of \textsc{model\_mags} does not impact the results strongly;][]{2014MNRAS.442.3443D}, matching to the ALLWISE catalog positions with a radius of 2 arcseconds. We limit the optical footprint to that of the BOSS survey\footnote{\url{http://www.sdss3.org/dr9/algorithms/boss_tiling.php\#footprint}}, which covers 10,269 deg$^2$.  The imaging in many of the ancillary regions outside of the BOSS footprint is less uniform, which complicates the splitting of the sample and can lead to problems with the relative obscured and unobscured density as a function of position.  We include \textsc{mangle} polygon files of the BOSS footprint, as well as a pixelated version of the DR8 footprint (which is also used in order to remove small regions where imaging is not complete) at online with the other mask files above.  We also apply the SDSS ``bad fields'' mask \citep[e.g.][]{2011ApJ...728..126W}, which is included online with the other masking files for completeness.

\begin{figure*}
\centering
\vspace{0.3cm}
\hspace{-2.5cm}
   \includegraphics[width=20cm]{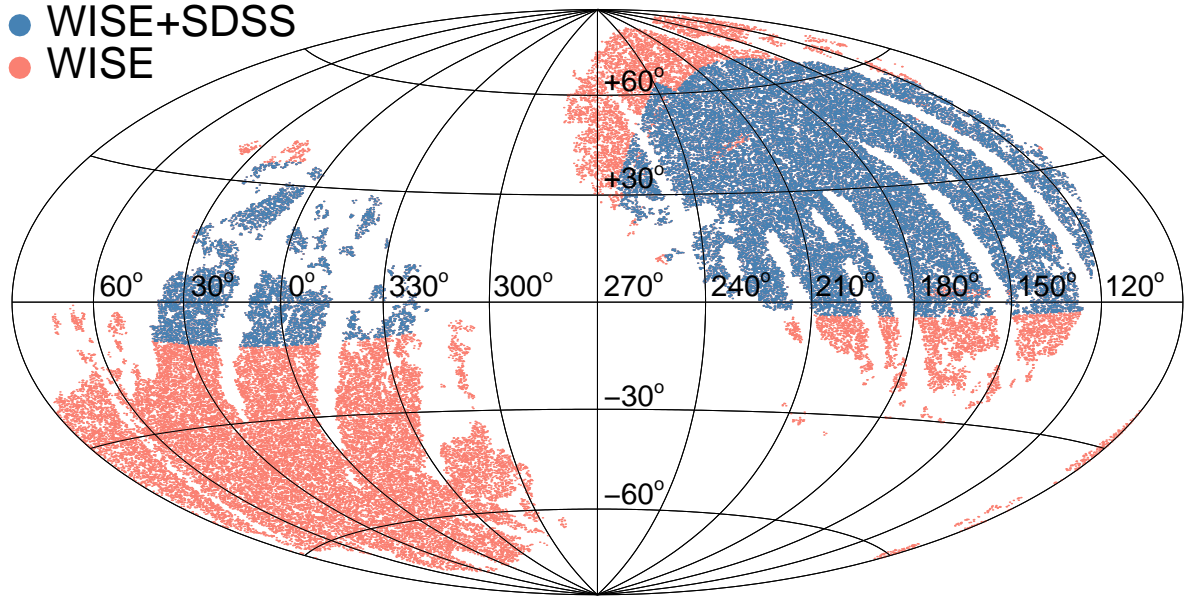}
    \vspace{-2cm}
  \caption{Aitoff projection (equatorial coordinates) of the quasars selected from \wise\ using the criteria of \citet{2012ApJ...753...30S}, after our mask is applied and the Galactic plane is removed (down-sampled to 15\% of the original samples).  The total sample of \wise\ quasars is shown in light red, and the objects overlapping the SDSS legacy footprint (where a split into obscured and unobscured is possible) are shown in blue.\label{fig:samples}}
\vspace{0.2cm}
\end{figure*}

Within the available SDSS$+$\wise\ footprint (5314 deg$^2$), there are 294,023 \wise-selected quasars, overlaid in blue on the full \wise-selected sample in red in Figure~\ref{fig:samples}.  Of these, 245,800 (83.5\%) have optical matches with $15 < r < 25$.  The $r-W2$ colours are calculated using the ALLWISE $W2$ magnitudes. Sources without optical detections are considered obscured, and the final samples consist of 172,537 (58.6\%) unobscured and 121,486 (41.4\%) obscured quasars, roughly a factor of two larger than the samples in D16. Due to more strict masking (Section~\ref{sec:lensing}), these numbers are reduced somewhat for the CMB lensing correlations.  We will consider the impacts of including or neglecting sources without optical matches in later sections.

\subsection{Redshift distributions and luminosities} \label{sec:redshifts}
\begin{figure}
\centering
   \includegraphics[width=8cm]{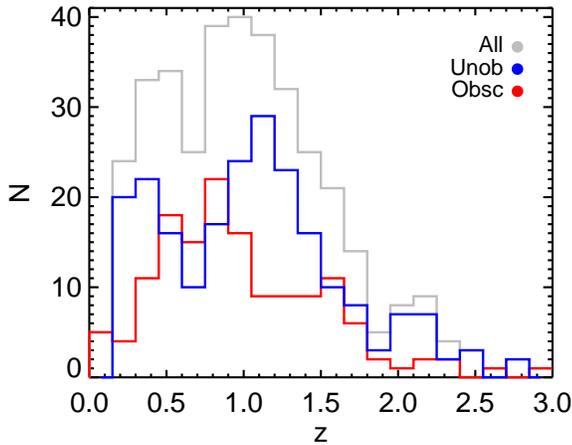}
  \caption{Redshift distributions of each sample, based on deep spectroscopy in the Bo\"{o}tes field. The mean of each is $\sim1$ (Table~\ref{tbl:results}).\label{fig:dndz}}
\end{figure}

In order for fair comparisons between the unobscured and obscured populations, they must be well-matched in redshift and luminosity.  However, without individual source redshifts, we are limited to characterizing the overall properties of the sample more generally by studying subsets in small fields with deep multi-band imaging and follow up spectroscopy. As in D16, we apply our selection criteria and mask to sources in the Bo\"{o}tes field \citep{2006ApJ...651..791B, 2007ApJ...671.1365H}, and find 368 \wise-selected quasars in the $\sim$7 deg$^2$ portion with deep spectroscopy \citep[][]{ 2012ApJS..200....8K}.  Nearly all of these sources (361, or 98\%) have spectroscopic redshifts, which allows us to generate accurate redshift distributions ($dN/dz$) for both the obscured and unobscured samples, shown in Figure~\ref{fig:dndz}.  Both samples cover a similar range ($0 < z < 3$) and have mean redshifts of $\sim$1 (0.98 and 1.05 for obscured and unobscured, respectively).  We will utilize cross-correlations between samples to further probe the similarity in redshift distributions in Section~\ref{sec:wise_xcorr}.

Because there are trends between black hole mass, galaxy mass, and halo mass \citep[e.g.][and references therein]{2004ApJ...604L..89H, 2010MNRAS.405L...1B, 2010MNRAS.404.1111G, 2013ARA&A..51..511K}, it is expected that there should also be correlations between quasar luminosity and halo mass.  However, observational evidence of this relationship has been elusive \citep[][]{2009ApJ...697.1656S, 2015MNRAS.453.2779E}, potentially due to the breadth of the Eddington ratio distribution and variability timescales \citep{2009ApJ...698.1550H, 2014ApJ...782....9H, 2016ApJ...826...12J}.  Regardless, because of this potential link, comparisons of obscured and unobscured quasar halo masses should only be made for similar luminosities. \citet{2007ApJ...671.1365H, 2011ApJ...731..117H} used detailed SED fitting and bolometric corrections to the \emph{Spitzer} rest-frame 8 $\mu$m luminosities of IR-selected quasars in the Bo\"{o}tes field to illustrate that simple IR colour-cuts select high luminosity quasars (peaking at $L_{\textrm{bol}} \sim 10^{46}$ erg/s), and that the obscured and unobscured populations selected using optical-IR colours have very similar luminosity distributions.

An additional complication in calculating intrinsic luminosities for \wise\ obscured quasars is that in the most sensitive (shorter wavelength) bands, the IR emission may still be significantly anisotropic \citep[e.g.][]{2011ApJ...736...26H}, as well as affected by reddening in sources with the highest column densities.  A simple check on the importance of these effects in our samples is to look at the detection fractions and magnitude distributions in the long-wavelength \wise\ bands, as well as colours across the \wise\ filters.  

\begin{figure}
\centering
   \includegraphics[width=8cm]{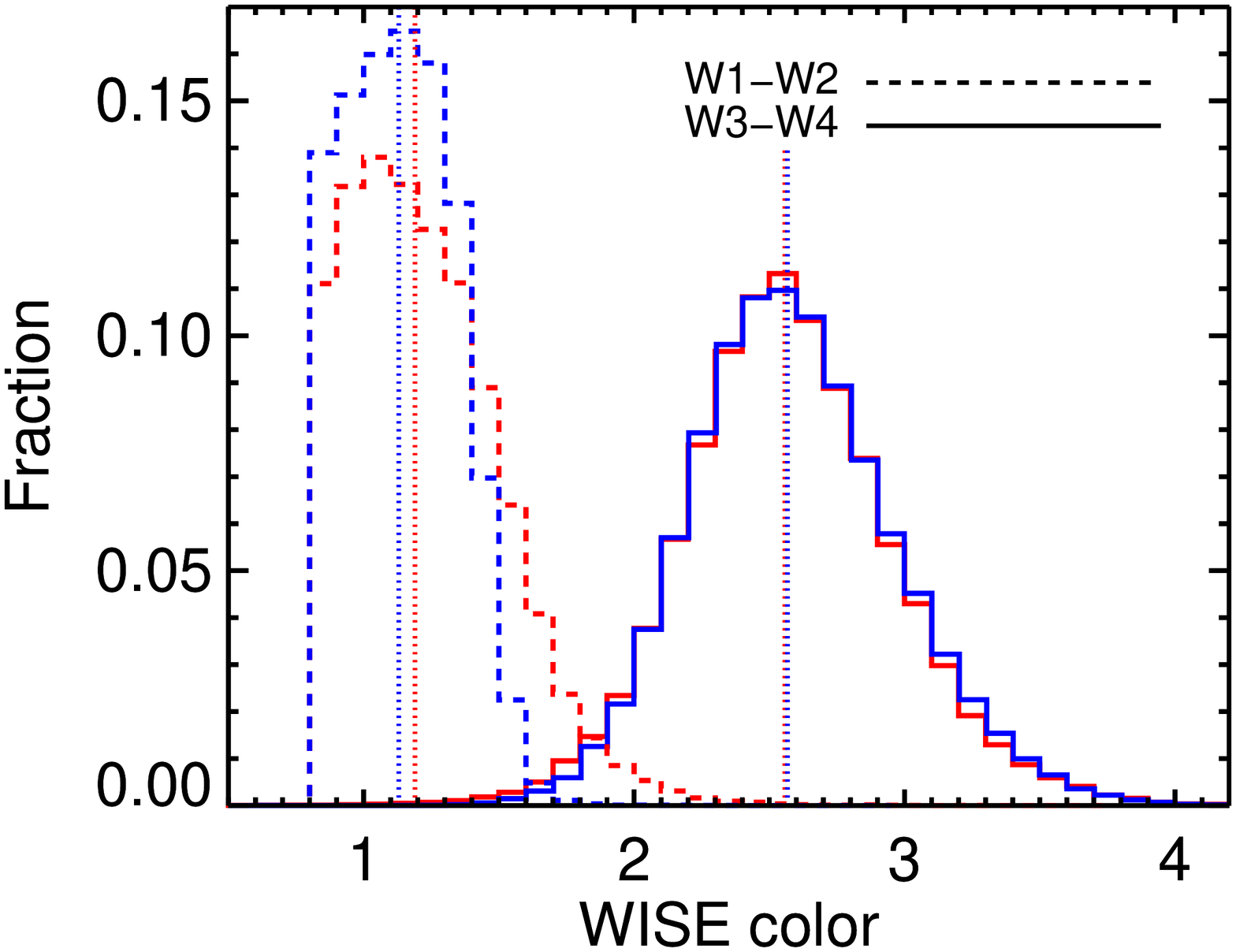}
   \includegraphics[width=8cm]{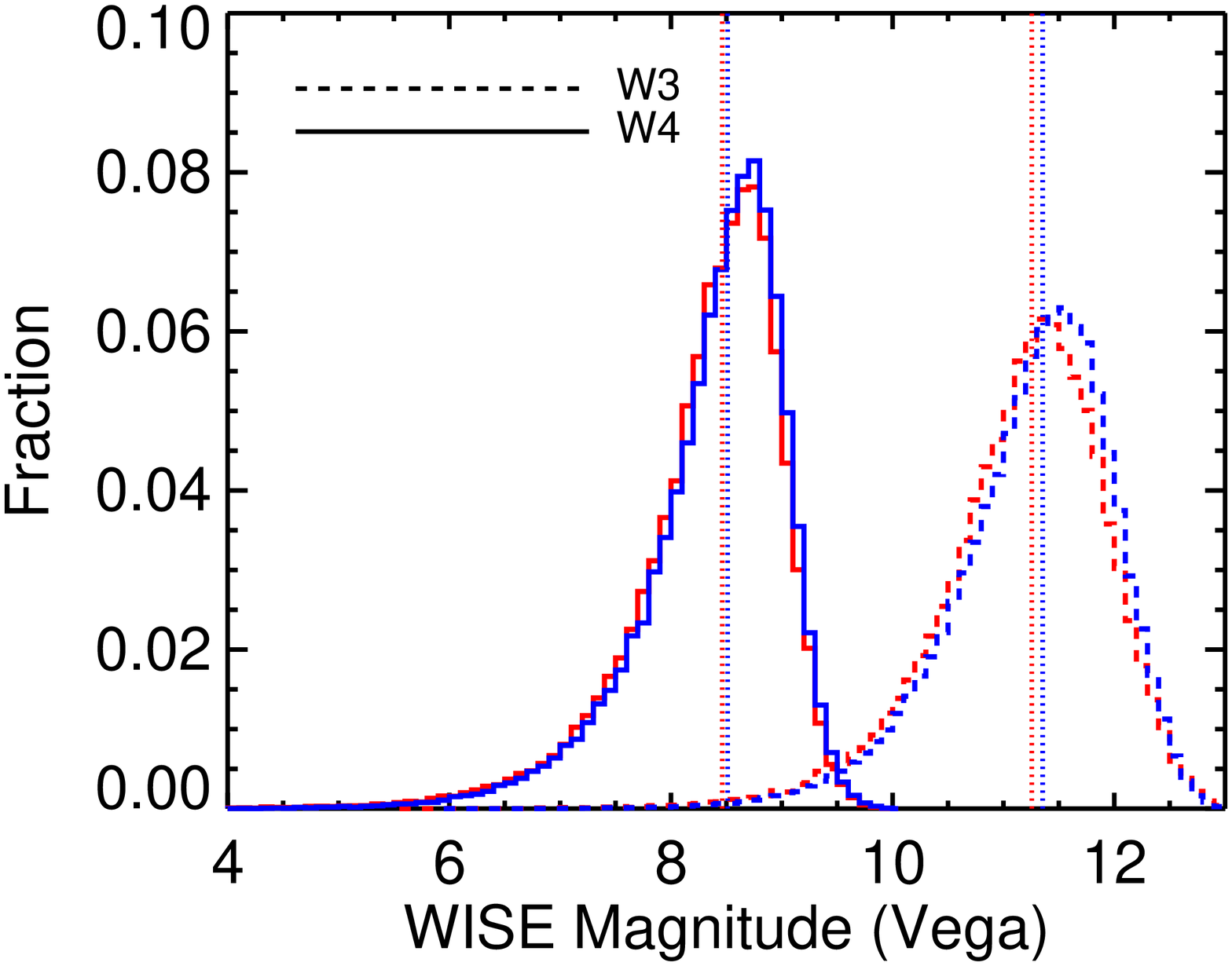}
  \caption{\emph{Top:} Colour distributions of the short wavelength (dashed) and long wavelength (solid) \wise\ bands for the obscured (red) and unobscured (blue) samples.  Since reddening is still a factor even in the mid-IR, a slight offset in obscured and unobscured $W1-W2$ colours is present, but because the long wavelength bands are not strongly affected the $W3-W4$ colour distributions are nearly identical. \emph{Bottom:} The magnitude distributions of the samples in the long-wavelength ($W3$ dashed and $W4$ solid, respectively) \wise\ bands.  Both samples have nearly identical median fluxes and detection fractions in these bands.\label{fig:w3w4}}
\end{figure}

By definition, all of our targets have $W1$ and $W2$ measurements. In the $W3$ and $W4$ bands, we find obscured detection fractions ($>5\sigma$) of 96\% and 64\%, respectively, while for the unobscured sample we find detection fractions of 97\%, and 62\%, respectively.  In the top panel of Figure~\ref{fig:w3w4} we show the colour distributions of each sample at shorter ($W1-W2$) and longer ($W3-W4$) wavelengths.  It is clear that at shorter wavelengths reddening (and potentially orientation) effects are present, as the obscured sample is redder on average due to a significant red tail in the distribution.  However, this difference completely disappears at the longer wavelengths, highlighting the utility of these bands for direct comparisons.

In the second panel of Figure~\ref{fig:w3w4} we show the magnitude distributions in $W3$ and $W4$ of the detected sources, and again find excellent agreement between the samples.  The median magnitudes in the detected obscured sources are 11.3 ($W3$) and  8.5 ($W4$), and 11.4 and 8.5 for the unobscured sources. Considering the well-matched redshift distributions, these similarities suggest excellent correspondence in luminosity.

\section{ANGULAR CLUSTERING} \label{sec:clustering}
\subsection{Methodology}
Without individual source redshifts, we are limited to projected angular correlation functions and rely on overall redshift distributions (see Section~\ref{sec:redshifts}).  D16 presented a detailed description of angular correlation functions, including the underlying theory, methods for calculation, and model generation\footnote{IDL codes used to produce our measurements are available at \url{https://github.com/mdipompe/angular_clustering}; see D16.}. We refer the reader to their section 3.1 for complete details, and provide a brief summary here.

The angular correlation function $\omega_{\theta}$ quantifies the probability that a pair of objects separated by angular distance $\theta$ and with mean number density $n$ will be found within the solid angle $d\Omega$ \citep{1969PASJ...21..221T, Peebles:1980vn}:
\begin{equation}
   dP = n(1+\omega_{\theta})d\Omega.
\end{equation}
Massive galaxies that host quasars are found in the largest potential wells of the dark matter distribution, and are therefore biased relative to the underlying density field \citep{1984ApJ...284L...9K, 1986ApJ...304...15B}.  We quantify this via the quasar bias $b_q$, which is tied to the underlying dark matter angular correlation function: $\omega_q(\theta) = \omega_{\textrm{dm}}(\theta) b_q^2$, under the limiting assumption that the bias is not a strong function of redshift or spatial scale.  In the case of a cross-correlation between two samples, rather than the autocorrelation of one, this becomes $\omega_{\textrm{cross}}(\theta) = \omega_{\textrm{dm}}(\theta) b_1 b_2$, where $b_1$ and $b_2$ are the biases of the two populations of interest (we will refer to the term $b_1b_2$ as $b_{\textrm{cross}}^2$ below).

To model $\omega_{\textrm{dm}}(\theta)$, we produce the non-linear matter power spectrum $P(k,z)$ using CAMB\footnote{\textit{Code for Anisotropies in the Microwave Background} (\url{http://lambda.gsfc.nasa.gov/toolbox/tb_camb_ov.cfm}), via our IDL wrapper \textsc{camb4idl} which can be found at \url{https://github.com/mdipompe/camb4idl}} \citep{2000ApJ...538..473L}.  On scales $<< 1$ radian, Limber's approximation can be used to project the matter power spectrum onto the sky \citep[in a flat Universe;][]{1953ApJ...117..134L, Peebles:1980vn, 1991MNRAS.253P...1P, 2007ApJ...658...85M}:
\begin{multline}
 \omega_{dm} (\theta) = \pi \int_{z=0}^{\infty} \int_{k=0}^{\infty} \frac{\Delta^2 (k,z)}{k} J_0 [k \theta \chi(z)] ~\times \\
    \left( \frac{dN}{dz} \right)_1 \left( \frac{dN}{dz} \right)_2 \left(\frac{dz}{d \chi} \right) \frac{dk}{k} dz.
\label{eq:omega_mod}
\end{multline}
Here, $\Delta^2 (k,z) = (k^3/2\pi^2)P(k,z)$ is the dimensionless power spectrum at redshift $z$ and wavenumber $k$, $J_0$ is the zeroth-order Bessel function of the first kind, $\chi$ is the comoving distance along the line of sight, and $dz/d\chi = H_z/c = (H_0/c)[\Omega_{\textrm{m}}(1+z)^3 + \Omega_{\Lambda}]^{1/2}$.  The $dN/dz$ terms represent the redshift distributions of the samples of interest (Section~\ref{sec:redshifts}), and in the case of an autocorrelation are identical. 

The actual angular autocorrelation measurements are made by comparing the number counts of quasar pairs on different scales with what is expected for a random distribution \citep{1993ApJ...412...64L}:
\begin{equation}
   \omega_{\textrm{qq}}(\theta) = \frac{DD - 2DR + RR}{RR},
   \label{eq:LS}
\end{equation}
where $DD$, $DR$, and $RR$ are the normalized numbers of data-data, data-random, and random-random pairs in each bin of $\theta$ (i.e. $DD=DD(\theta)=N_{\textrm{data pairs}}/(N_D N_D)$). In the case of a cross-correlation, the $2DR$ term is split into $D_1R_2 + D_2 R_1$ and the $RR$ terms become $R_1 R_2$, where the subscripts indicate the distinct samples in their corresponding random catalogs. 

The random data must have the same angular distribution as the real data for an accurate comparison in the presence of selection effects.  In previous work over a limited footprint the quasar samples were consistent with being uniformly distributed over the field (after masking), simplifying the generation of the random catalog using \textsc{mangle} \citep{2004MNRAS.349..115H, 2008MNRAS.387.1391S}.  In the expanded footprint used here, there are position-dependent fluctuations in the quasar density, and we apply corrections to an initially uniform random sample (see Section~\ref{sec:allwise}).  In all cases, the final random catalog is always at least 50 times larger than the real data set so that Poisson noise in the random counts do not limit the precision of the measurement.  In general, the angular correlation function is calculated in four bins per dex, from $\sim$7 arcseconds to 1.1 degrees, unless otherwise noted. 

We use inverse-variance weighted jackknife resampling to estimate uncertainty on the angular correlations \citep[e.g.][D16]{2002ApJ...579...48S, 2005MNRAS.359..741M, 2007ApJ...658...85M}, with the footprint broken into $N=100$ equal-area regions.  We iteratively remove one region at a time and repeat the measurement in the remaining area.  This provides the covariance matrix $\textrm{\textsf{\textbf{C}}}_{ij} = \textrm{\textsf{\textbf{C}}}(\theta_i,\theta_j)$ ($i$ and $j$ denote angular size bins):
\begin{multline}
\textrm{\textsf{\textbf{C}}}_{ij} = \sum_{L=1}^{N} \sqrt{\frac{RR_L(\theta_i)}{RR(\theta_i)}} [\omega_L(\theta_i) - \omega(\theta_i)] ~\times  \\
      \sqrt{\frac{RR_L(\theta_j)}{RR(\theta_j)}} [\omega_L(\theta_j) - \omega(\theta_j)],
\label{eq:jack}
\end{multline}
where $\omega$ is the angular correlation function for the entire sample and $\omega_L$ is the same for subset $L$.  Here, the $RR$ terms are not normalized by the size of the random sample, allowing the ratio to compensate for the different number counts in each region in the case where the areas are not precisely equal.  The square-root of the diagonal elements of the covariance matrix are adopted as the 1-$\sigma$ uncertainty on the measurements.

In this work we introduce an updated fitting technique beyond the simple $\chi^2$ minimization used in prior work.  The motivation for this update is that the uncertainties on these types of clustering measurements are not necessarily Gaussian, which can impact the results of the fit.  We note that in the case of our measurements here, the distribution of $\omega(\theta)$ values at each angular scale are roughly Gaussian --- however, we introduce this fitting procedure here in case it is useful to the community and so that our procedure is more general for future work.  It also allows us to generate a posterior distribution of bias values, which may not be purely Gaussian and therefore not fully parametrized with a single value and error bar.

The jackknife resampling provides 100 measurements of $\omega(\theta)$ at each angular scale.  We draw randomly from these values to create a new autocorrelation measurement, and assign Poisson errors to each based on the number of pairs contributing to each scale:
\begin{equation}
\sigma_p^2(\theta) = \frac{f(1+\omega(\theta))^2}{DD}
\end{equation}
where $DD$ is the unnormalized number of pairs, and $f=2$ in the case of an autocorrelation (because pairs are not independent) and $f=1$ in the case of a cross-correlation.  We then fit our dark matter autocorrelation model using standard $\chi^2$ minimization, and repeat the process $n=10,000$ times to build a distribution of bias values.  When single values and uncertainties are reported, they represent the median and symmetric 67\% confidence interval of this distribution\footnote{We note that this technique is a pseudo-Bayesian approach.  We have also implemented a fully Bayesian MCMC approach \citep[based on e.g.][]{Hogg:2010p1751} --- however fully incorporating non-Gaussian uncertainties into this method is computationally expensive, whereas the method used here produces very similar results while still allowing for arbitrary uncertainty distributions.}.

\begin{table*}
  \caption{Summary of samples and bias and halo mass measurements.}
  \label{tbl:results}
  \begin{tabular}{lccccc}
  \hline
       Sample                   &  &         N        & $\langle z \rangle$ &          $b_q$             & $\log(M_h/M_{\odot} \ h^{-1}) $    \\
  \hline
                                      &  &                            \multicolumn{4}{c}{Angular Clustering}                \\          
                                                                                      \cline{2-6}       \\               
All                                  &  &   581,728    &           1.02             &    1.98$\pm$0.05     & $12.79_{-0.04}^{+0.04}$ \\
All (SDSS footprint)       &  &   296,793    &           1.02             &     1.94$\pm$0.07    & $12.75_{-0.06}^{+0.06}$ \\
Unobscured                   &  &   172,537    &           1.05             &    1.77$\pm$0.11     & $12.56_{-0.12}^{+0.11}$ \\
Obscured                       &  &   121,486    &           0.98             &    2.15$\pm$0.12     & $12.96_{-0.09}^{+0.08}$ \\
\\
                                       &  &                            \multicolumn{4}{c}{CMB Lensing Cross-correlation}                \\          
                                                                                      \cline{2-6}   \\                   
All                                  &  &   531,545    &           1.02             &   1.82$\pm$0.14       &    $12.65_{-0.15}^{+0.13}$          \\
All (SDSS)                     &  &   269,751    &           1.02             &    1.85$\pm$0.19      &     $12.68_{-0.19}^{+0.16}$         \\
Unobscured                   &  &   157,172   &           1.05             &    1.66$\pm$0.23      &      $12.45_{-0.31}^{+0.23}$        \\
Obscured                       &  &   94,938     &           0.98             &   2.14$\pm$0.27      &       $12.95_{-0.22}^{+0.18}$       \\
\\
                                       &  &                            \multicolumn{4}{c}{Combined$^a$}                \\          
                                                                                      \cline{2-6}   \\          
Unobscured                   &  &                   &          1.05              &    1.70$\pm$0.10   &   $12.49_{-0.11}^{+0.10}$   \\
Obscured                       &  &                   &          0.98              &    2.15$\pm$0.11   &   $12.94_{-0.08}^{+0.08}$   \\
\hline
\multicolumn{6}{l}{$^a$ Weighted mean of clustering and lensing results, after verifying sufficient independence; see Section~\ref{sec:discussion}.}
   \end{tabular}
   \\  
{
\raggedright    
 }
\end{table*}

\subsection{Results}
\subsubsection{All \wise\ Quasars} \label{sec:allwise}
\begin{figure}
\centering
\vspace{0cm}
\hspace{0cm}
   \includegraphics[angle=90,width=8.5cm]{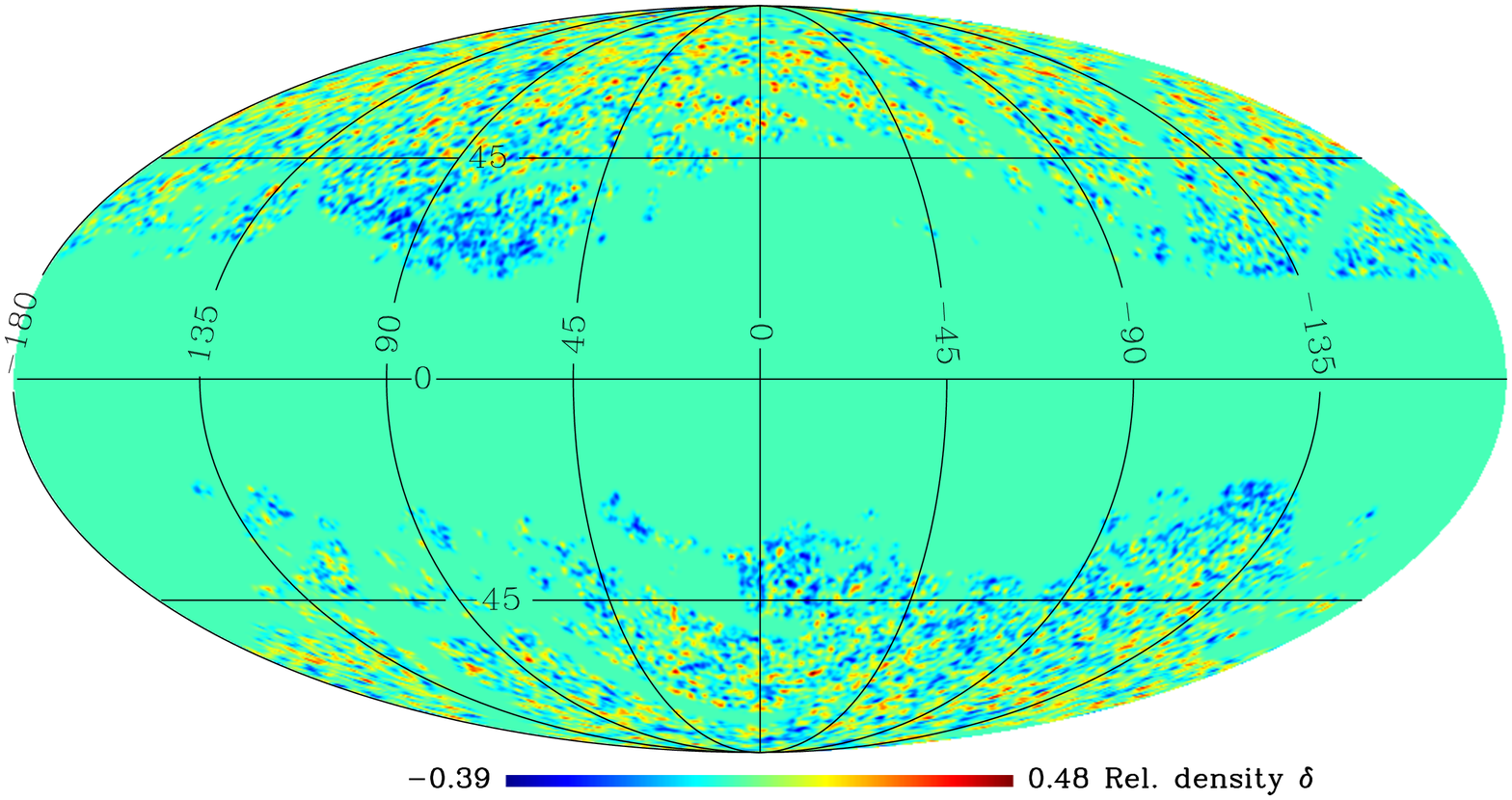}
   \includegraphics[width=8cm]{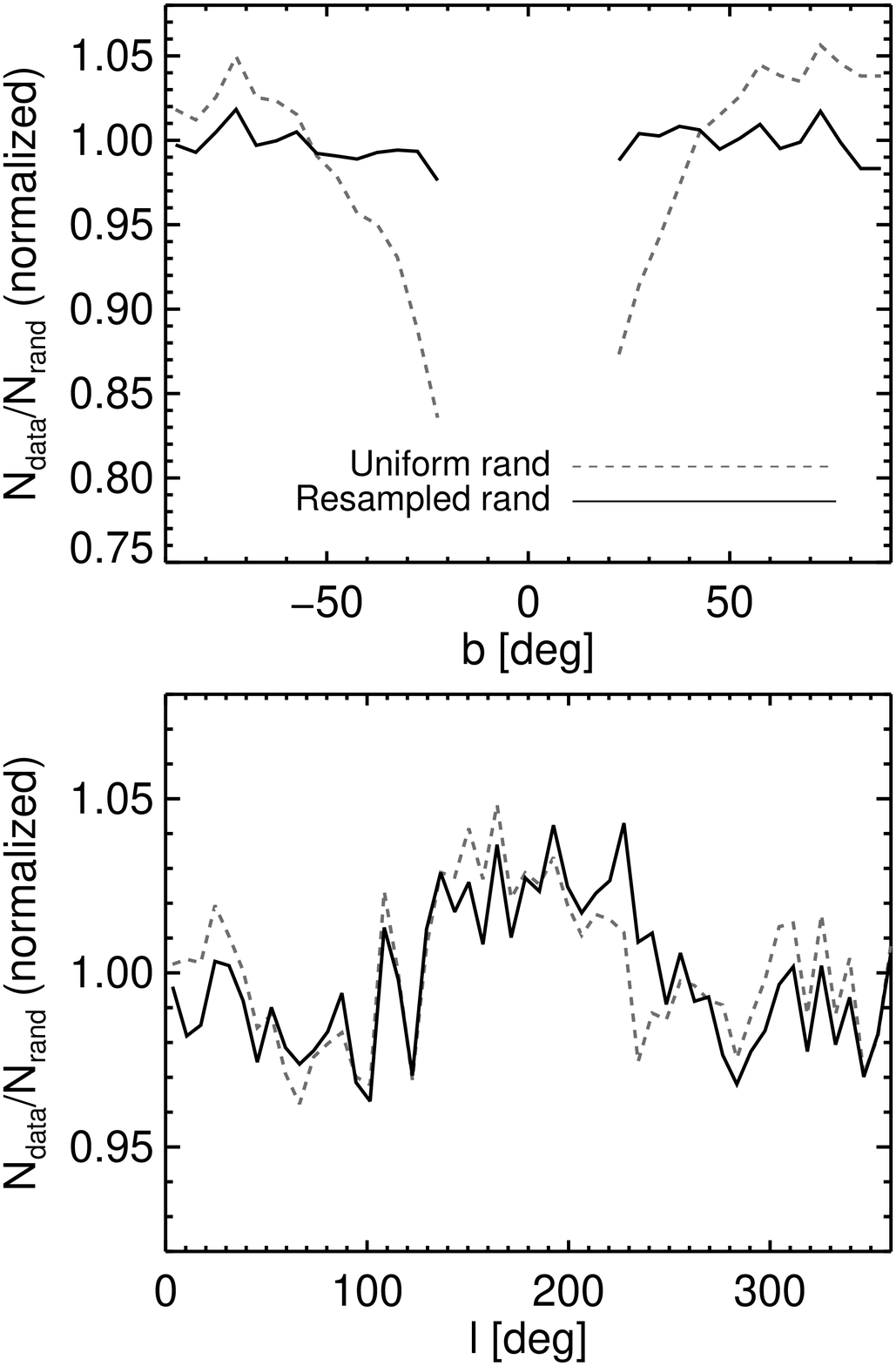}
    \vspace{0cm}
  \caption{\emph{Top:} Mollweide projection (Galactic coordinates) of the density of \wise\ quasars relative to the overall mean density ($\delta_q$), smoothed with a 1 degree Gaussian. There is a clear decrease in density at lower Galactic latitude, likely due to confusion with the increasing stellar density.  \emph{Centre and bottom:} Taking the ratios of normalized data and random distributions in Galactic $b$ (centre) and $l$ (bottom) highlights that a uniform random distribution (dashed grey) does not reflect the data, with the strongest variation in $b$.  We correct the random distribution in the $b$ direction (solid black), as described in Section~\ref{sec:allwise}.\label{fig:rand_dist}}
\vspace{0cm}
\end{figure}

We first measure the angular autocorrelation of the largest quasar sample to date --- the 581,728 selected with \wise\ from the entire sky (after masking).  However, expanding beyond the initial pilot region of D16 presents a complication in that the quasar density is not uniform across the sky, as shown in the top panel of Figure~\ref{fig:rand_dist}.  There is a decrease in the number of quasars selected closer to the Galactic plane, even after removing $|b| < 20$ degrees, likely because the stellar density increases and it becomes more likely that a Galactic star falls within a \wise\ resolution element of a distant quasar \citep[this was also noted in e.g.][]{2015ApJ...803....4K}.  In the lower two panels of Figure~\ref{fig:rand_dist}, we normalize histograms of the Galactic coordinates of the quasar sample and a uniformly distributed random sample and take their ratio (dashed lines) --- it is clear that the systematic variation is with Galactic latitude, while variations in Galactic longitude appear mostly dominated by noise.

In order to correct the random sample such that it follows the data distribution, we first pixelate the sky into equal area regions using HEALPix \citep{2005ApJ...622..759G}, with $n_{\textrm{side}} = 8$, which has $\sim$7 deg$^2$ pixels.  We explored different pixel sizes and settled on the smallest pixel size possible without obvious ``over-fitting'' such that intrinsic variance was being traced.  We calculate the quasar density in each pixel (taking care to trace the available area of each pixel in the presence of our mask), and take the mean in pixels with the same central latitude.  Normalizing these to a maximum value of unity provides weights for each pixel as a function of $b$, which are then used to remove an appropriate number of random sources from each pixel.  This provides a resampled random catalog that better matches the distribution of the data, as can be seen in the solid lines of the lower panels of Figure~\ref{fig:rand_dist}.

\begin{figure}
\centering
\vspace{0cm}
\hspace{0cm}
   \includegraphics[width=7.9cm]{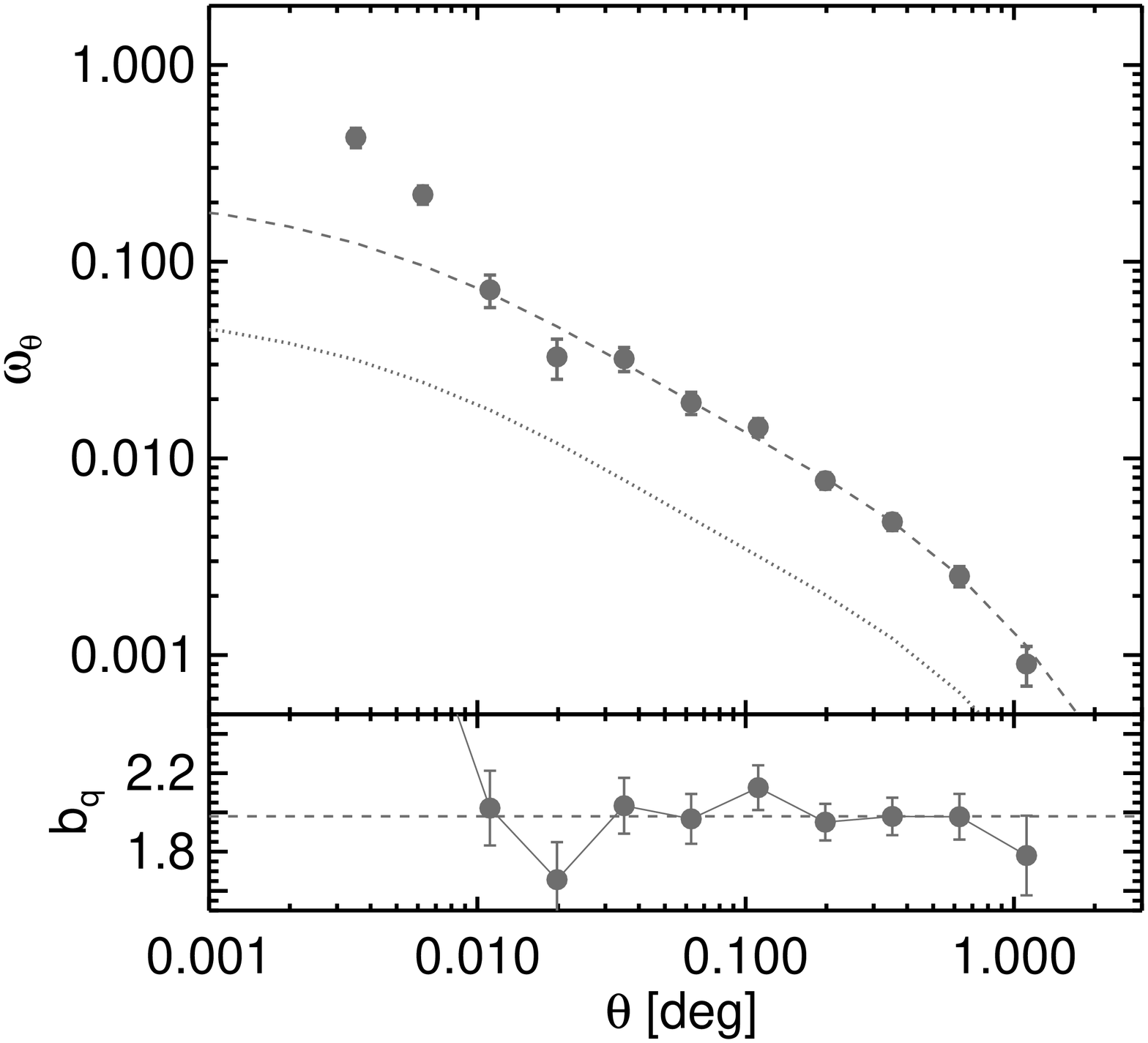}
   \includegraphics[width=7.9cm]{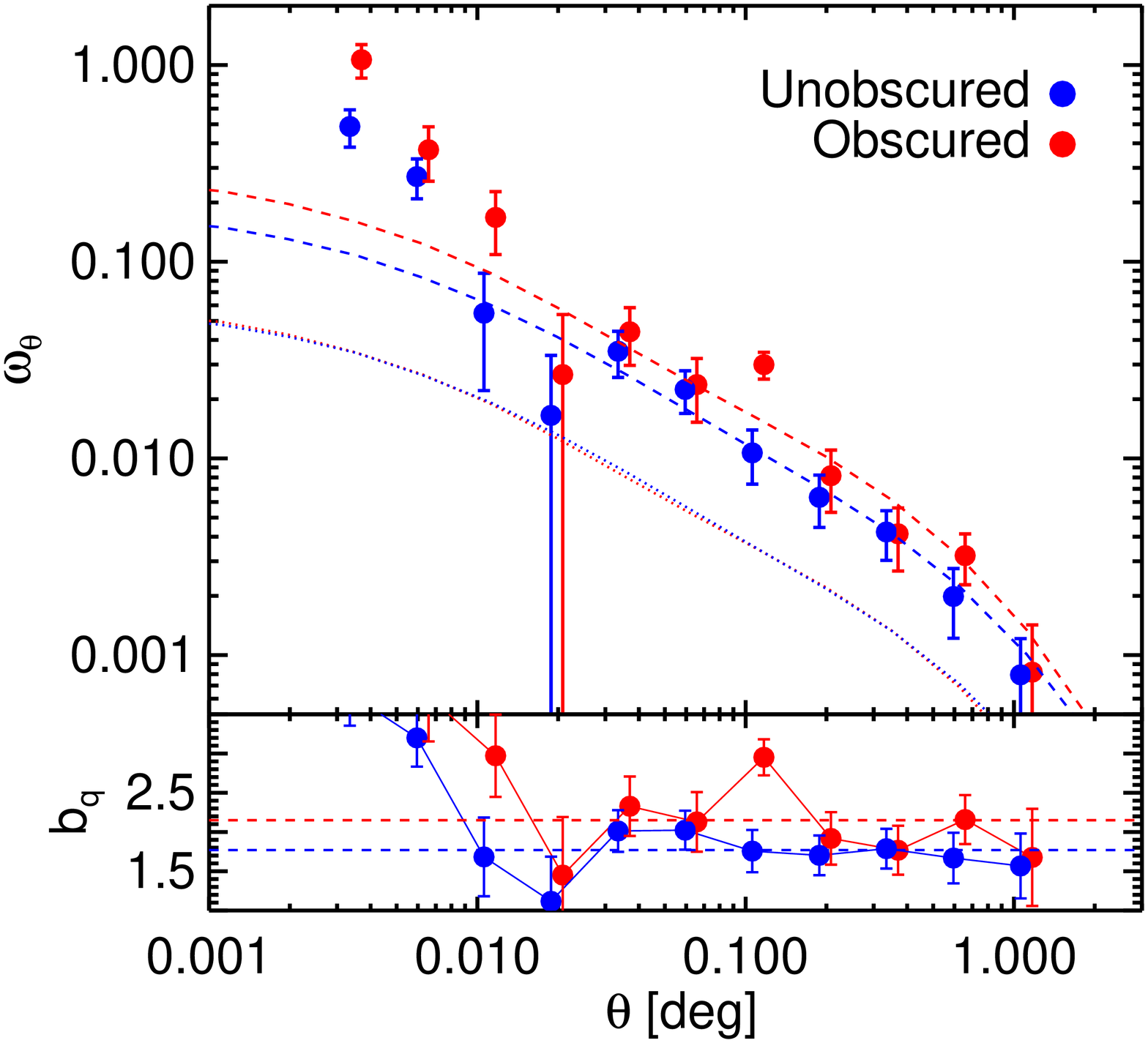}

    \hspace{0.4cm}   \includegraphics[width=7.5cm]{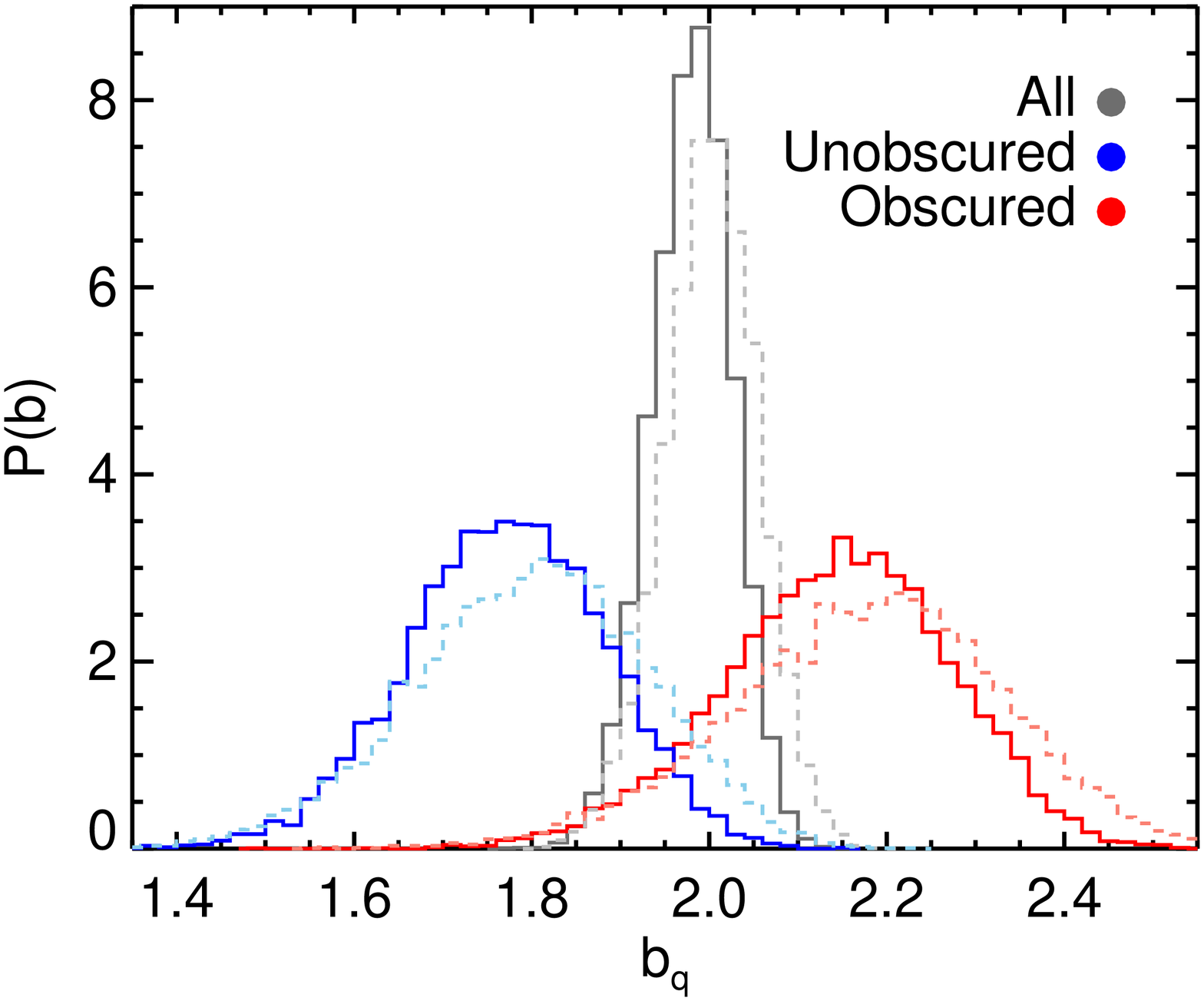}
    \vspace{0cm}
  \caption{\emph{Top:} The angular autocorrelation of all 581,728 \wise-selected quasars. The dashed line shows the dark matter autocorrelation model (dotted line) rescaled by $b_q^2$ to fit the data.  The second panel shows the square root of the data divided by the model, another way to visualize $b_q$. \emph{Middle:} The angular autocorrelation of the obscured (red) and unobscured (blue) \wise-selected quasars in the SDSS footprint. Dotted lines show the model dark matter autocorrelations, and the dashed lines show the model re-scaled to fit the data through the bias. \emph{Bottom:} Bias posterior distributions. Dark, solid lines are based on fits over the wider 0.025 to 1.5 degree range, and lighter, dashed lines are fits over the more limited 0.04 to 0.4 degree range.\label{fig:cluster_wise}}
\vspace{0cm}
\end{figure}

The angular autocorrelation for the full \wise-selected sample is shown in the top panel of Figure~\ref{fig:cluster_wise}.  With such a large sample, we find an excellent clustering signal with error bars smaller than the data points in many cases.  The data match the shape of the model remarkably well on scales where the two-halo term dominates (where pairs are largely quasars in distinct haloes rather than subhaloes, on scales of a few tenths of a Mpc --- note that at $z=1$, 0.01 degrees corresponds to $\sim$0.4 Mpc/$h$ comoving), including the turnover on scales approaching one degree.  On small scales, there appears to be significant signal from the one-halo term, which can be used to probe the quasar halo occupation distribution and satellite fraction \citep[though see the discussion in the next section regarding these scales for the obscured and unobscured populations;][]{2005ApJ...633..791Z, 2007ApJ...659....1Z, 2013ApJ...779..147C}.  

Fitting the dark matter model to these data (on scales $> 0.025$ degrees, which will be adopted throughout unless otherwise noted) results in a \wise-selected quasar bias of $1.98\pm0.05$ (Table~\ref{tbl:results}).

\subsubsection{Obscured and Unobscured Quasars} \label{sec:obsc_unob}
We now turn to the angular autocorrelation of \wise\ quasars split into obscured and unobscured subtypes --- these results are shown in the middle panel of Figure~\ref{fig:cluster_wise}. We apply a similar correction to the random sample in the SDSS footprint as for the full \wise\ sample, though the fact that the SDSS footprint does not extend as far in to the Galactic plane means that the correction is not as significant.  Again, we overall see excellent agreement with the shape of the models, including the turnover on large scales.  

Fitting the models to these data results in $b_{\textrm{unob}} = 1.77 \pm 0.11$ and $b_{\textrm{obsc}} = 2.15 \pm 0.12$ (Table~\ref{tbl:results}). These are consistent with D16, but the reduced uncertainties due to the increase in sample size means that this difference is $\sim3.3\sigma$. Using our new procedure for fitting the bias, we also extract posterior probability distributions for the bias of each sample, shown in the lower panel of Figure~\ref{fig:cluster_wise}. We note that these distributions are not purely Gaussian, and tend to have a slight asymmetry with a larger tail toward lower bias values.  There is only a small amount of overlap between the obscured and unobscured distributions --- integrating the obscured PDF where it overlaps the unobscured distribution yields a total overlap probability of 0.08, while the unobscured overlap with the obscured distribution is 0.05.

The obscured autocorrelation does show an anomalous increase on scales of $\sim 0.1$ degrees, with this point falling roughly 3$\sigma$ from the best-fit line. While this could just be a statistical fluctuation, as a check that this is not due to contaminants that specifically affect this scale (e.g.\ haloes around bright stars, globular clusters, etc.), we visually inspected 100 random SDSS images of objects contributing to this angular bin, as well as another larger bin for comparison.  While this is a small fraction of the roughly 100,000 pairs that contribute to this bin, we found no evidence of additional contaminants on this scale.  To test whether this single point is a strong driver of the enhanced obscured bias, we re-perform the fit with this scale excluded. This results in an obscured bias reduced slightly to $b_{\textrm{obsc}} = 2.06 \pm 0.16$, which is consistent with the full measurement and still larger than the unobscured bias.

Additional signal on small scales, again possibly due to the one-halo term and providing valuable information on the satellite fraction (e.g. Mitra et al.\ in prep), is present in both of these subsamples as well.  However, we do note that in D14 it was found that after additional masking of the \wise\ quasar sample \citep[as compared to][]{2014ApJ...789...44D} the autocorrelation signal was reduced most prominently in the obscured sample on small scales.  This suggests that we exercise caution when analyzing these scales, as they are sensitive to artifacts in the data.  The fact that the unobscured sample shows this excess signal, and was not affected strongly by the additional masking in D14, however suggests that it may be real.  In addition, D16 found that incorporating the updated ALLWISE data release, which we also use here, removed a significant amount of contamination, based on comparison of the positional dependence of the quasar density and the signal on increasingly large scales.

It is possible that the non-optically detected subset of the obscured sample (roughly 40\% of the full obscured sample) suffers more contamination and affects our measurement of the bias.  We repeat the autocorrelation measurement using only the optically-detected obscured sample, and find that they are nearly identical.  The increased signal at 0.1 degrees is largely smoothed over, and the resulting bias measurement is $b_{\textrm{obsc,opt}} = 2.18 \pm 0.19$.

\subsubsection{\wise\ Obscured-Unobscured Cross-Correlation} \label{sec:wise_xcorr}
\begin{figure}
\centering
\vspace{0cm}
\hspace{0cm}
   \includegraphics[width=8cm]{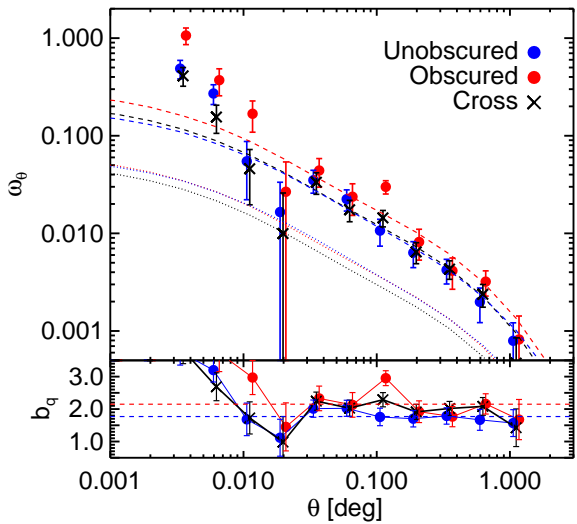}
    \vspace{0cm}
  \caption{Cross-correlation of the obscured and unobscured quasars (crosses), compared to the autocorrelations.  Note that the cross-correlation model (grey dotted line) is lower in amplitude than the autocorrelation models (red and blue dotted lines), which results in slightly larger biases inferred from the cross-correlation. However, the results are consistent and the cross-correlation suggests a good match between the samples in redshift space.\label{fig:cross}}
\vspace{0cm}
\end{figure}

Because we are only able to estimate the redshift distributions of the samples based on data in the Bo\"{o}tes field, it is informative to explore the cross-correlation of the obscured and unobscured populations, which carries extra information about the overlap of the samples in redshift space. We show the cross-correlation compared to the autocorrelations in Figure~\ref{fig:cross}.  Qualitatively, the cross-correlation behaves as we would expect for samples well-matched in redshift, with the clustering amplitude falling generally in between the individual autocorrelations.

As the model curves in Figure~\ref{fig:cross} (dotted lines) show, the cross-correlation model is slightly lower in amplitude than the autocorrelation models due to small differences in the redshift distributions (Figure~\ref{fig:dndz}).  Thus, when fitting the model to the cross-correlation, a slightly larger rescaling is needed.  Since the cross-correlation signal is proportional to $b_{\textrm{obsc}} b_{\textrm{unob}}$, we can fit the model to the data, and divide by the biases inferred from the autocorrelations to determine the bias of one sample or the other.  These results are listed in Table~\ref{tbl:cross}, and while the decreased model amplitude for the cross-correlations produces slightly larger bias values for each sample, these results are quite consistent with the results from the autocorrelation alone.

\begin{table}
  \caption{Summary of cross-correlation results}
  \label{tbl:cross}
  \begin{tabular}{lcccc}
  \hline 
                                                                    &  &    $b_1 b_2$         &  $b_{\textrm{unob}}$ & $b_{\textrm{obsc}}$      \\
  \hline
Obsc $\times$ Unob                 &  &   4.12$\pm$0.41    &      1.92$\pm$0.22    &    2.33$\pm$0.27      \\
Spec $\times$ Unob                &  &   3.06$\pm$1.26    &      1.1$\pm$0.5        &    ...                       \\
Spec $\times$ Obsc                 &  &   5.29$\pm$1.38    &      ...                         &    1.9$\pm$0.6      \\
\hline
   \end{tabular}
   \\  
{
\raggedright    
 }
\end{table}

If we assume that the unobscured redshift distribution is better constrained than that of the obscured sample, which is not unreasonable given the optical brightness of the unobscured objects, we can use the cross-correlation and models generated using simulated obscured redshift distributions to inform us about the true obscured $dN/dz$.  To this end, we generate Gaussian $dN/dz$ distributions with standard deviation 0.55 (similar to the standard deviation from sources in the Bo\"{o}tes field) and means $\mu_2$ from 0.1 to 2 in steps of 0.1.  Note that negative values of $z$ are not possible, so these are discarded and the positive portions of the distributions are properly re-normalized such that $\int \frac{dN}{dz}dz =1$ in all cases.

\begin{figure}
\centering
\vspace{0cm}
\hspace{0cm}
   \includegraphics[width=8cm]{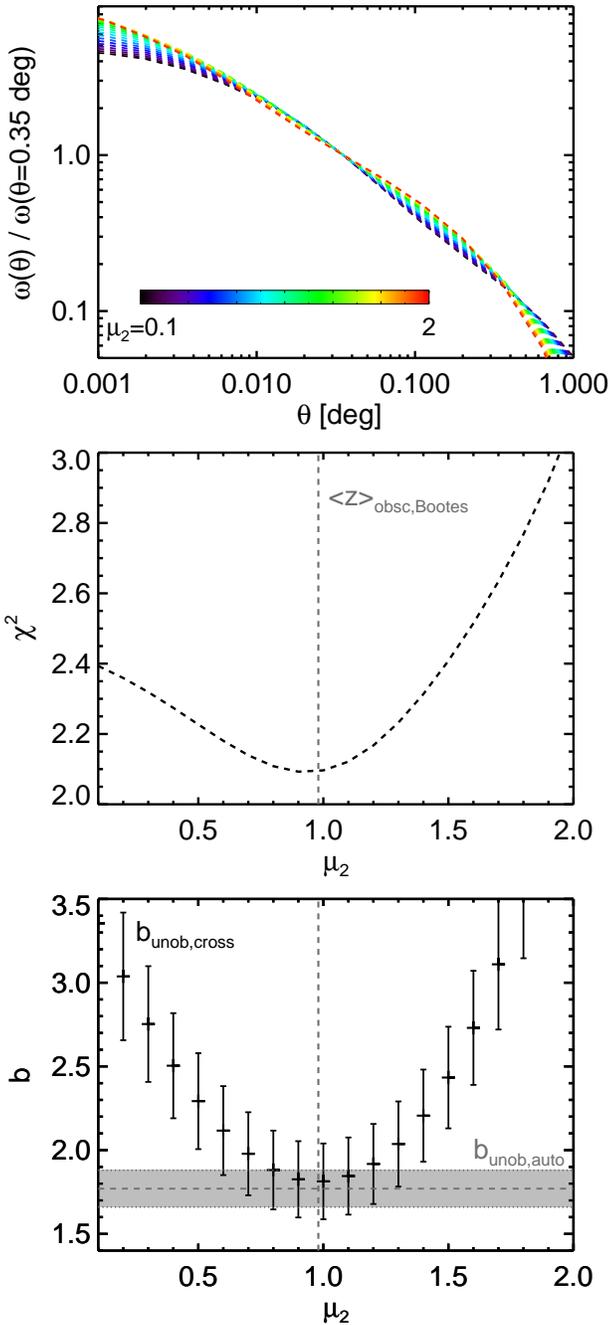}
    \vspace{0cm}
  \caption{\emph{Top:} Model obscured-unobscured cross-correlations using Gaussians for the obscured $dN/dz$ (and the actual unobscured $dN/dz$) with standard deviation 0.55 and mean $\mu_2$.  These are normalized at 0.035 degrees to highlight differences in the shapes of the models. \emph{Middle:} The models are fit to the cross-correlation (Figure~\ref{fig:cross}), and we show the resulting $\chi^2$ for each $\mu_2$. There is a clear minimum at $\mu_2 \sim 1$, in excellent agreement with the mean obscured redshift of objects in the Bo\"{o}tes field (vertical dashed line). \emph{Bottom:} The unobscured bias derived from the cross-correlation and the models using mock redshift distributions, via Equation~\ref{eq:model}, along with the unobscured bias from the autocorrelation (grey region).  A mean obscured redshift of 1 provides the best agreement between the unobscured autocorrelation and the cross-correlation.\label{fig:cross_models}}
\vspace{0cm}
\end{figure}

In the top panel of Figure~\ref{fig:cross_models}, we show model dark matter cross-correlations using the observed unobscured $dN/dz$ and the mock obscured distributions, normalized to unity at $\sim$0.035 degrees.  Note that the amplitudes of the models vary, but we show the normalized versions here to highlight that there is also variation in the shapes depending on the $z$ overlap.  

In the second panel of Figure~\ref{fig:cross_models} we show the $\chi^2$ values obtained by fitting each of these models to the cross-correlation, which provides information on how well the \textit{shape} (not amplitude) of the data matches the various models.  There is a clear minimum when $\mu_2 \sim 1$, suggesting that the obscured sample is not at significantly different mean redshift than the analysis of the Bo\"{o}tes field implies.

Taking this one step farther, and assuming that the unobscured autocorrelation (and resulting bias) is robust, we can use the above models to estimate the unobscured bias from the cross-correlation for each $\mu_2$, and search for the model that provides the best agreement with the unobscured autocorrelation.  In addition to the model cross-correlations, we also generate model obscured autocorrelations for each $\mu_2$.  We fit these model autocorrelations to the obscured data, and fit the corresponding model cross-correlation to the cross-correlation data, which provides the unobscured bias as a function of $\mu_2$:
\begin{equation}
\label{eq:model}
b_{\textrm{unob}}=\frac{b_{\textrm{cross}}^2 (\mu_2)}{b_{\textrm{obsc}} (\mu_2)}.
\end{equation}
We compare these results with the unobscured bias derived from the unobscured autocorrelation alone in the last panel of Figure~\ref{fig:cross_models}.  Again, we see that the best agreement from the two approaches is when the mean redshift of the obscured sample is at $\sim$1. This is additional confirmation that the obscured $dN/dz$ we infer from the Bo\"{o}tes field is not significantly skewed by observational biases.

\begin{figure}
\centering
\vspace{0cm}
\hspace{0cm}
   \includegraphics[width=8cm]{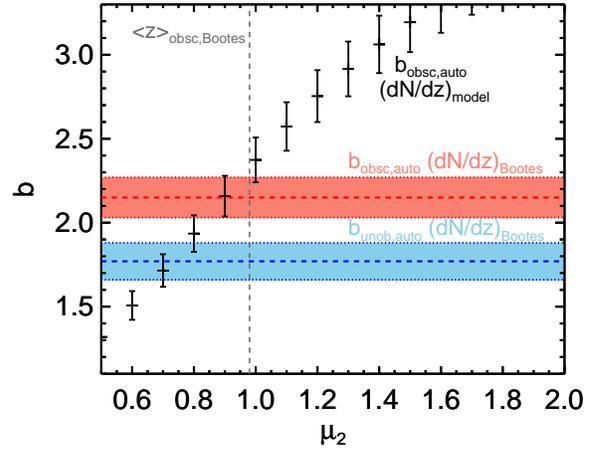}
    \vspace{0cm}
  \caption{The obscured bias inferred using models with mock Gaussian $z$ distributions with mean $\mu_2$, used to determine the unobscured bias from the cross-correlation in the lower panel of Figure~\ref{fig:cross_models}. The biases (with uncertainties) from the autocorrelations are shown as horizontal lines/bands --- if the mean obscured redshift is lower than estimated, it could become consistent with the unobscured bias, but the bottom panels of Figure~\ref{fig:cross_models} suggest that this is not the case (see Section~\ref{sec:wise_xcorr}).\label{fig:obsc_b_z}}
\vspace{0cm}
\end{figure}

Figure~\ref{fig:obsc_b_z} shows the obscured bias inferred from the autocorrelation with each mock $dN/dz$, with the values from the autocorrelations using the Bo\"{o}tes redshift distributions overlaid.  In order for the obscured bias to be consistent with that of the unobscured sample, the mean redshift would have to be less than $\sim$0.8, which seems unlikely based on the observational data and the modeling of the cross-correlations in Figure~\ref{fig:cross_models}.  To further quantify this, we explored artificially inflating the obscured $dN/dz$ values (Figure~\ref{fig:dndz}) at $z<0.8$ until the overall mean was less than 0.8 --- in order for the true obscured mean to be this low, we would need to be missing roughly half of the population at low redshift.

Finally, we note that we have also performed these analyses using the obscured redshift distribution inferred from the Bo\"{o}tes field and shifting its mean while keeping the same shape, rather than using mock Gaussians.  The overall conclusions do not change.

\subsubsection{Cross-correlation with SDSS Spectroscopic Quasars} \label{sec:sdss_xcorr}
\begin{figure}
\centering
\vspace{0cm}
\hspace{0cm}
   \includegraphics[width=8cm]{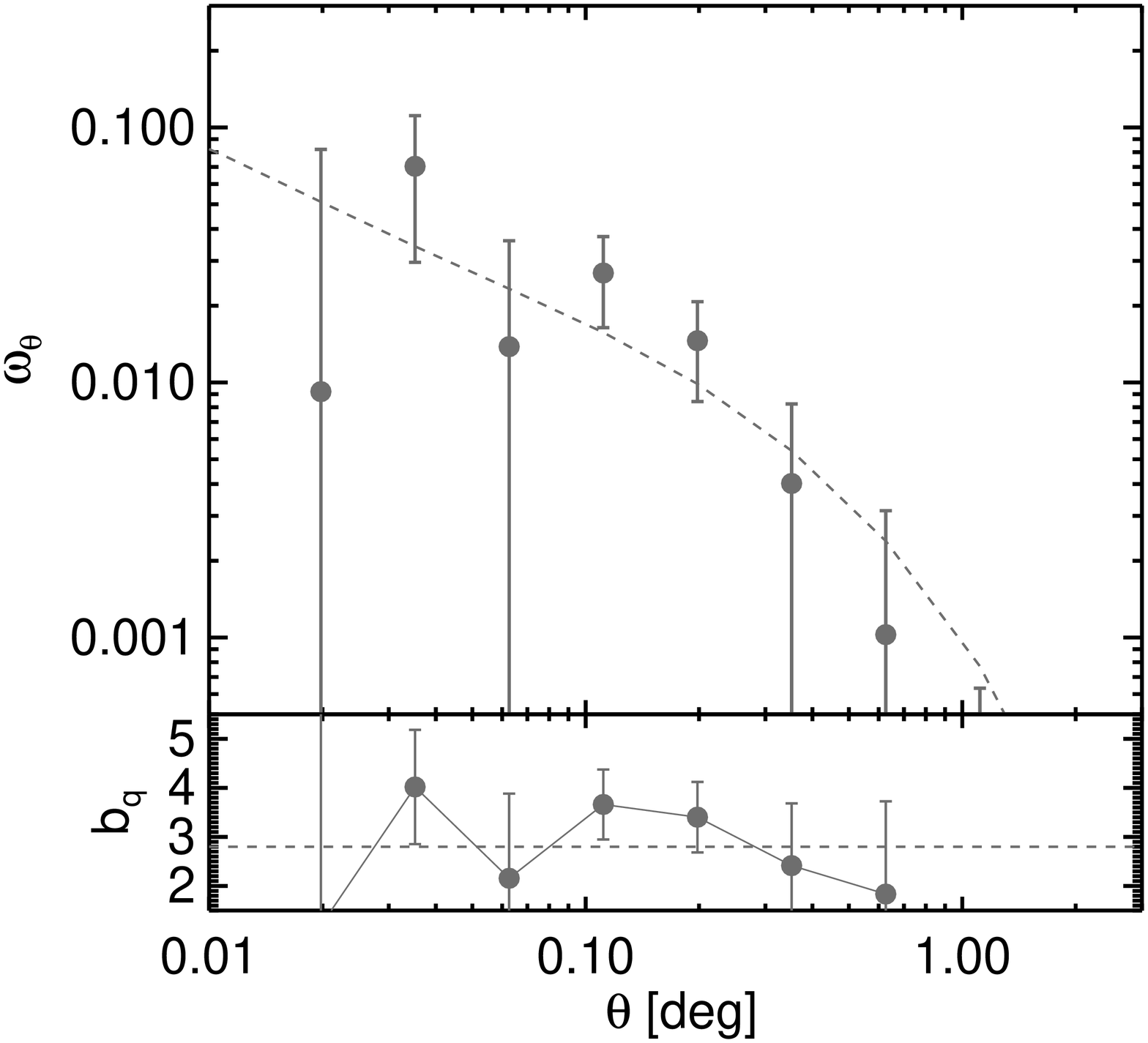}
   \includegraphics[width=8cm]{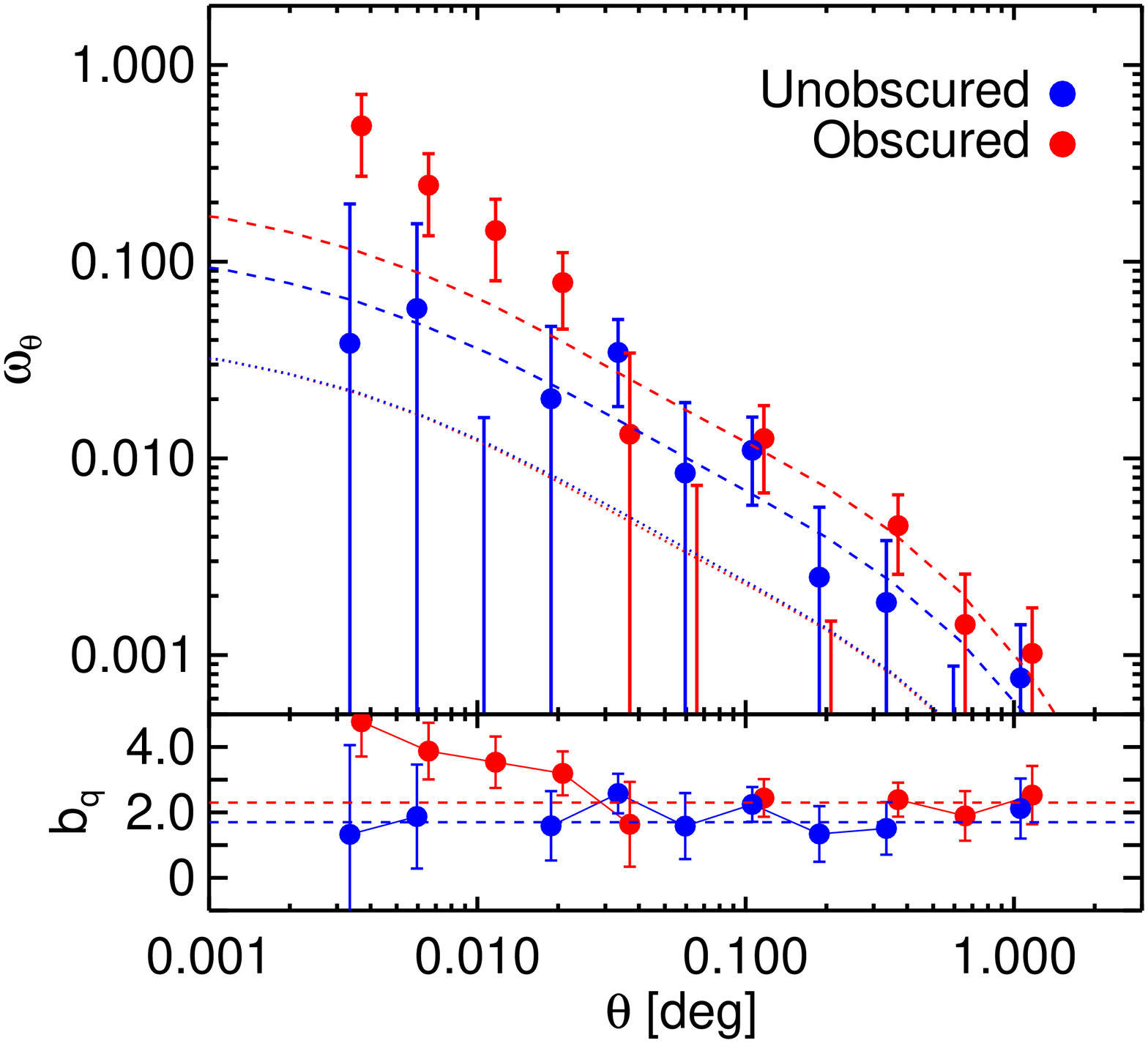}
    \vspace{0cm}
  \caption{\emph{Top:} Angular autocorrelation of SDSS spectroscopic quasars, with a measured bias consistent with that of \citet{2009ApJ...697.1656S}. \emph{Bottom:} Cross-correlation of SDSS spectroscopic and \wise\ quasar subsamples.  While noise is increased due to the low density of the spectroscopic sample, we find general agreement with the results of the autocorrelations (Tables~\ref{tbl:results} and~\ref{tbl:cross}).\label{fig:sdss_cross}}
\vspace{0cm}
\end{figure}

In order to study the obscured and unobscured biases in an independent way, we also cross-correlate each \wise\ sample with spectroscopically confirmed quasars from the SDSS.  We begin with the sample of \citet{2009ApJ...697.1656S}, which has a well-studied bias and complementary random catalogue that mimics the data distribution (Shen, private communication).  We first confirm that we can reproduce the bias results of \citet{2009ApJ...697.1656S} with an angular measurement, as the original work was done in three dimensions since the spectroscopic sample has individual source redshifts.  This is shown in the top panel of Figure~\ref{fig:sdss_cross}.  Despite the additional scatter in the angular measurement, we find good agreement with \citet{2009ApJ...697.1656S} --- our inferred bias is $b_{\textrm{spec}} = 2.7 \pm 0.5$, compared to $\sim2.4 \pm 0.3$.  

We limit the spectroscopic sample to the footprint available for the \wise\ data, discarding the Southern Galactic Cap where there is very little overlap between the samples. We also only include sources with $z_{\textrm{spec}} < 3$, where there is overlap with the \wise\ quasars.  We confirm that we find a consistent bias using this subset ($b_{\textrm{spec}} = 2.8 \pm 0.6$), and we adopt this value for the remainder of the analysis.

The cross-correlations between the spectroscopic and \wise-selected obscured and unobscured quasars are shown in the lower panel of Figure~\ref{fig:sdss_cross}.  Since the density of spectroscopic quasars is fairly low, the cross-correlation has larger uncertainty than the autocorrelations.  However, fitting models to these data and using the spectroscopic autocorrelation results, we find bias values lower than but generally consistent with those of the autocorrelations, as shown in Table~\ref{tbl:cross}.  

\section{CMB Lensing Correlations} \label{sec:lensing}
\subsection{Methodology} \label{sec:lensing_theory}
As CMB photons travel to us from the surface of last scattering at $z \sim 1100$, they are deflected by the matter distribution along the line of sight.  By cross-correlating maps of the quasar density with maps of the CMB lensing convergence $\kappa$, the quasar bias $b_q$ can be determined in yet another independent way.  As with the angular correlation functions, the details of the theory behind these measurements \citep{2012ApJ...753L...9B, 2012PhRvD..86h3006S}, the calculations, and generation of models are discussed in detail in D16.  We refer the reader there for complete details, and provide a basic summary here\footnote{See also our code library for these measurements at \url{https://github.com/mdipompe/lensing_xcorr}.}.

For a quasar sample with bias $b_q$ (which may be a function of redshift and therefore distance) their host dark matter halo distribution kernel is:
\begin{equation} 
W^q(\chi) = \frac{dz}{d\chi}\frac{dN}{dz} b_q(\chi),
\end{equation}
where $\chi$ is the comoving distance.  The CMB lensing kernel, which describes the strength of lensing as a function of distance to the lens, since the CMB is at a fixed distance, is given by \citep{2000ApJ...534..533C, 2003ApJ...590..664S}:
\begin{equation} W^\kappa(\chi) = \frac{3}{2}\Omega_{\rm
m}\left(\frac{H_0}{c}\right)^2\frac{\chi}{a(\chi)}\frac{\chi_{\rm CMB} - \chi}{\chi_{\rm CMB}},
\end{equation}
where $a(\chi)$ is the scale factor and $\chi_{\textrm{CMB}}$ is the co-moving distance to the CMB.  The cross-correlation in Fourier space at mode $\ell$ of these two components is given by: 
\begin{equation} C^{\kappa q}_l = \int dz
\frac{d\chi}{dz}\frac{1}{\chi^2}W^\kappa(\chi)W^q(\chi)P\left(\frac{l}{\chi},z
\right),
\label{eq:lens_model}
\end{equation}
where $P(k=\ell/\chi,z)$ is the matter power spectrum, again generated using CAMB.  Setting the bias in the distribution kernel equal to unity provides the model cross-power spectrum for the underlying dark matter distribution.

We utilize the all-sky CMB lensing potential map from the \emph{Planck} second data release \citep{2016A&A...594A...9P, 2016A&A...594A..15P} to perform the cross-correlation.  Using tools within \textsc{HEALpix}, the native format of the \emph{Planck} data, we convert the provided coefficients ($A_{\ell m}$) of the spherical harmonic transform of the lensing potential $\phi$ into an $n_{\textrm{side}} = 2048$ (resolution of $\sim$1.7 arcmin) map of the lensing convergence ($\kappa = 1/2 \nabla^2 \phi$).

The quasar positions are converted into a \textsc{HEALPix} map (with the same resolution) of the relative quasar density:
\begin{equation}
\delta_q = \frac{\rho - \bar{\rho}}{\bar{\rho}},
\end{equation}
where $\rho$ is the density in a given \textsc{HEALPix} pixel and $\bar{\rho}$ is the mean density.  To account for the fact that mean density changes as a function of position (see the top panel of Figure~\ref{fig:rand_dist}), rather than use the overall mean density over the full footprint, the relative density is calculated with respect to the local mean density in larger ($n_{\textrm{side}}=8$ or $\sim$7 deg) pixels.

Finally, we take the Fourier transform of each map and multiply them, keeping only the real portion, using the \textsc{HEALpix} routine \textsc{anafast}. We bin the cross-power in $\ell$, using 4 bins per dex, as with the angular correlation functions. Uncertainties on the cross-correlation are obtained by repeating the measurement with several rotated \planck\ lensing maps, which should have null cross-correlation signal in the absence of noise and systematics.  We use 17 rotations in Galactic longitude in steps of 20 degrees, and 17 more with an additional 180 degree rotation in latitude, for a total of 34 rotated maps \citep[see][and D16 for more details on this procedure to generate uncertainties]{2015MNRAS.446.3492D}. Cross-correlating with these maps provides the covariance matrix:
\begin{equation}
\textsf{\textbf{C}}_{ij} = \frac{1}{N-1} \left[ \sum_{k=1}^{N} (C_{l_i,k}^{\kappa q} - C_{l_i}^{\kappa q}) (C_{l_j,k}^{\kappa q} - C_{l_j}^{\kappa q}) \right],
\end{equation}
where the terms in parentheses are the cross-correlations with a given map rotation $k$ and $N=34$.

To fit the models generated with equation~\ref{eq:lens_model} to the data, we use a similar approach as for the clustering measurements.  We draw randomly a value of $C_{l}$ for each $\ell$ from the cross-correlations with the rotated maps, and add it to the actual measurement at that $\ell$.  We then use typical $\chi^2$ minimization to fit the model to this iteration, using the full covariance matrix in the fit (we are not able to generate Poisson errors as in the clustering measurements because it is not possible to separate out how many quasars contribute to the measurement at a given $\ell$).  Repeating the process $n=10,000$ times allows us to build a posterior distribution of bias values.

\subsection{Results} \label{sec:lensing_results}
While we defer a detailed discussion to Section~\ref{sec:discussion}, we have included an additional cut on the data for the CMB lensing cross-correlation, limiting our analysis to Galactic latitude $|b| > 40$ degrees. This is due to apparent increasing systematic effects with the expanded footprint that results in no significant improvement in the precision, as well as systematically lower bias values, particularly in the unobscured sample ($b_{\textrm{unob}} =1.5$, quite low for quasars selected in any way). Our ``best'' final result could be considered to be the result with both the best accuracy and the best precision. Precision might be expected to be driven by the total number of objects included in our analysis. Accuracy, however, will be a strong function of the systematics that contaminate our measurements. Our main sources of systematics (likely Moon levels, stellar density, Galactic dust, increasing contamination, etc.) tend to be most prevalent at low Galactic latitudes. Making this additional cut, while reducing the number of objects contributing to the measurement, leads to similar precision as the full footprint (and still slightly worse than D16).  We note that the clustering results are not strongly impacted by applying a similar cut.

\begin{figure}
\centering
\vspace{0cm}
\hspace{0cm}
   \includegraphics[width=8cm]{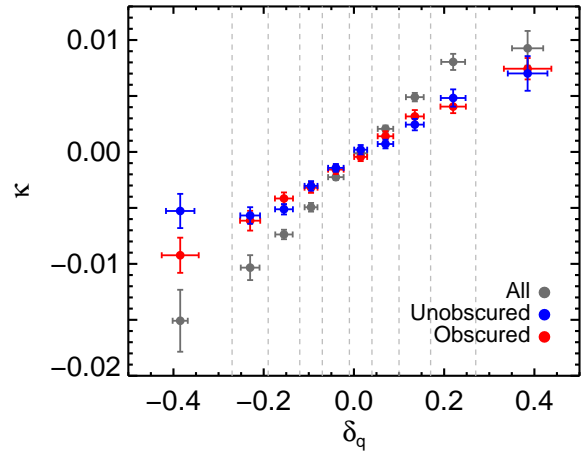}
    \vspace{0cm}
  \caption{Simple representation of the correlation between the quasar density $\delta_q$ and CMB lensing convergence $\kappa$. The maps are smoothed with a 1 degree Gaussian, and then binned in $\delta_q$, with bin edges shown by the grey dashed lines.  The mean value of $\kappa$ is then calculated in the pixels contributing to each bin.  Horizontal error bars are the 1$\sigma$ scatter in $\delta_q$ in each bin, and vertical error bars are the 1$\sigma$ scatter in $\kappa$ values when the density maps are cross-correlated with the rotated lensing maps to estimate uncertainties in the full cross-correlation analysis (see Section~\ref{sec:lensing}).\label{fig:lens_stack}}
\vspace{0cm}
\end{figure}

Figure~\ref{fig:lens_stack} highlights the high level of correlation between the quasar density maps and CMB lensing convergence.  We identify pixels within bins of $\delta_q$, and calculate the mean value of $\kappa$ in these pixels --- clearly, as the quasar relative density increases, the lensing convergence increases as well, showing that the quasar host haloes are indeed lensing the CMB photons.  Differences in the shape of this relationship for different samples will be discussed in Section~\ref{sec:discussion}.

\begin{figure}
\centering
\vspace{0cm}
\hspace{0cm}
   \includegraphics[width=8cm]{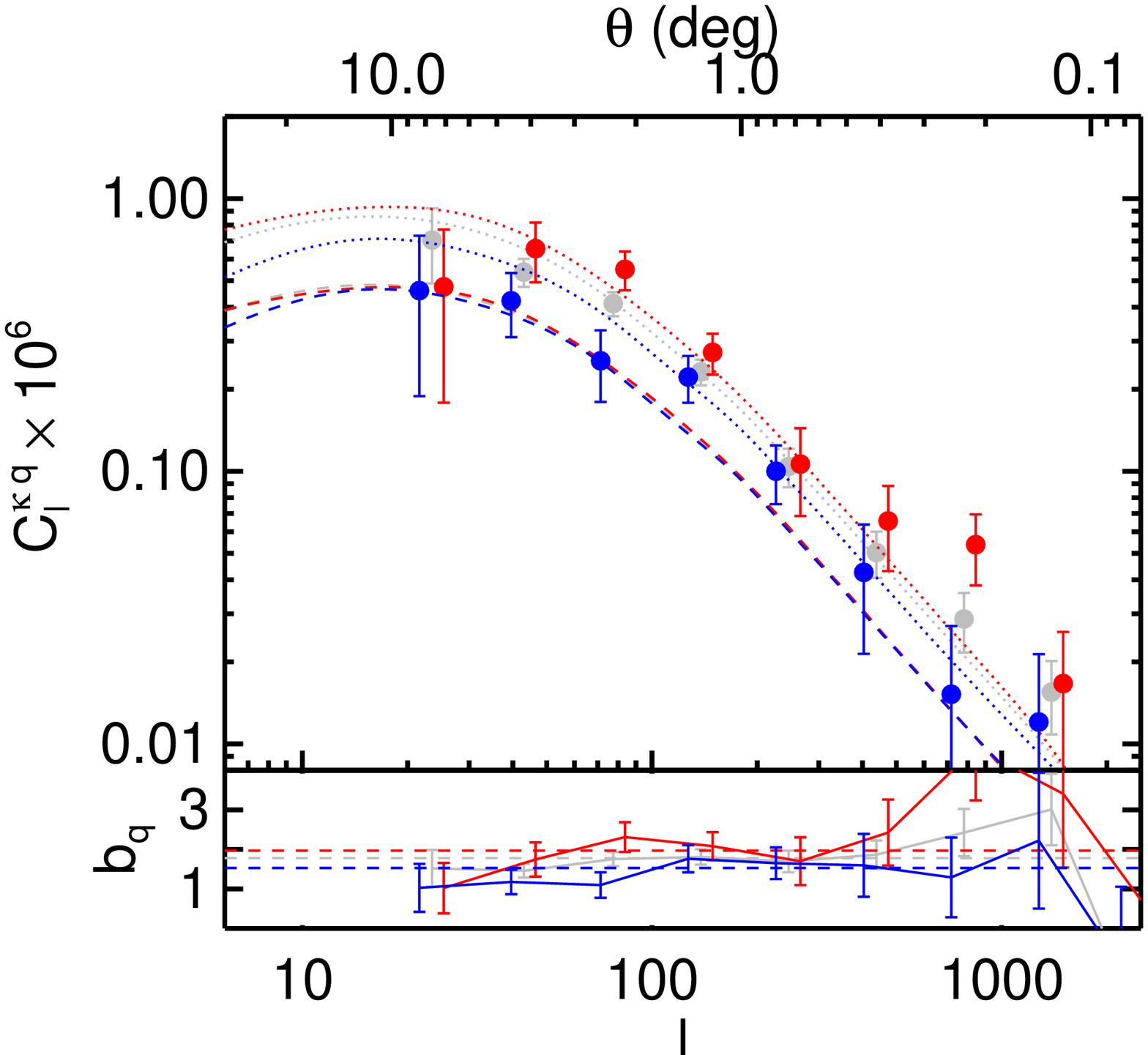}
   \includegraphics[width=8cm]{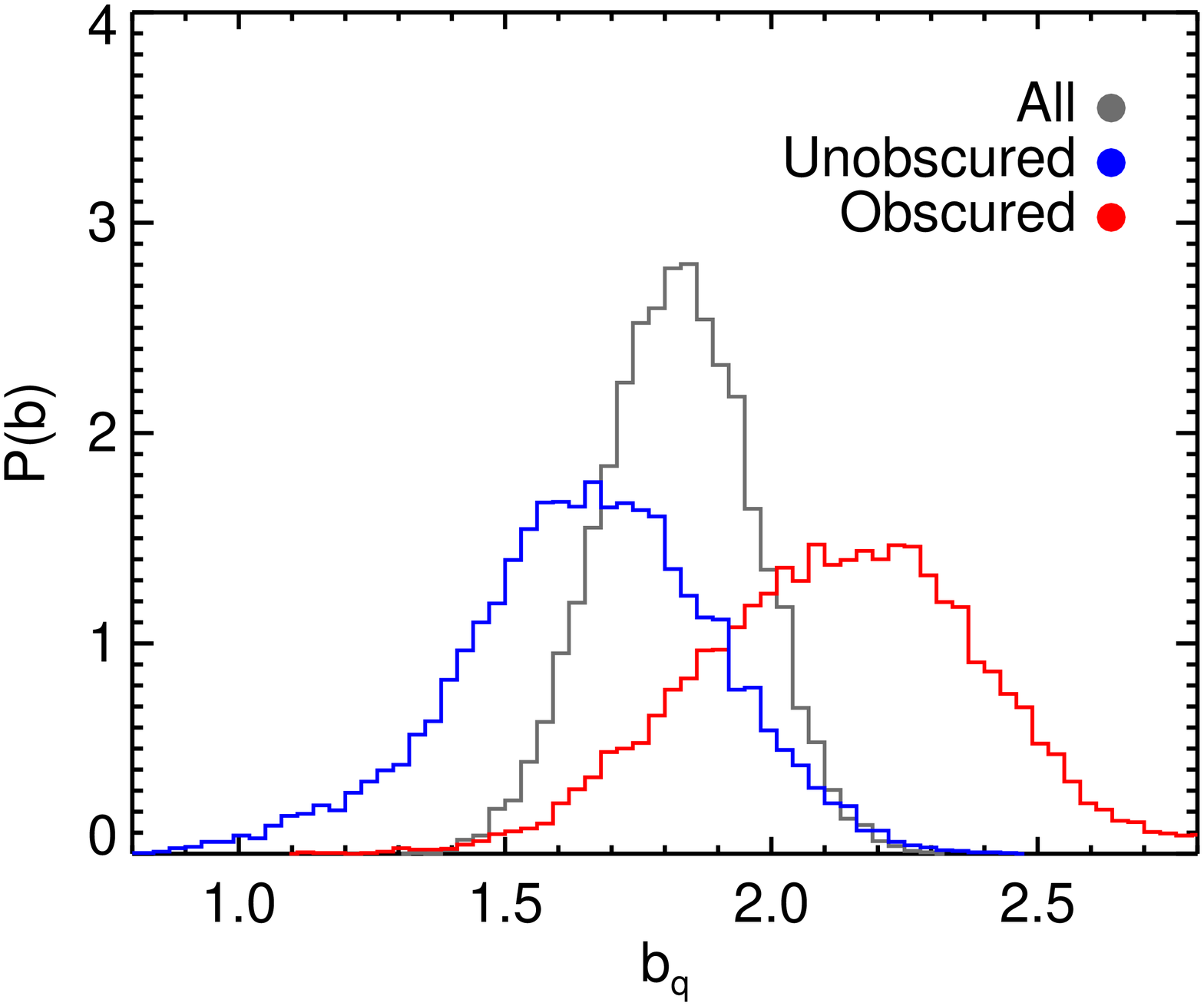}
    \vspace{0cm}
  \caption{\emph{Top:} Cross-correlation of \wise\ quasars over the full available footprint shown in grey, and for unobscured and obscured quasars in the SDSS footprint shown in blue and red, respectively.  The dashed lines show the model cross-correlation for the underlying dark matter distribution, and the dotted lines show the models fit to the data. The second panel shows the data divided by the models in order to better visualize the bias (the full sample is not shown here for clarity).  \emph{Bottom:} Bias posterior distributions from fits to the CMB lensing cross-correlations. These are wider than the distributions from clustering, but the obscured and unobscured quasars still have distinct peaks. The medians and confidence intervals from these distributions are given in Table~\ref{tbl:results}.\label{fig:lensing}}
\vspace{0cm}
\end{figure}

The formal quasar and CMB lensing cross-correlations in this high Galactic latitude measurement are shown in the top panel of Figure~\ref{fig:lensing}, where the grey points show the results for the complete \wise\ sample not limited to the SDSS footprint, and the blue and red points are the unobscured and obscured results over the more limited footprint, respectively.  A few important things are evident in this figure.  First, the obscured quasar cross-correlation signal is higher on nearly all scales.  Second, there is an increase in the obscured cross-correlation signal on scales of $\sim$0.2 degrees, which is similar to the scales at which additional signal was present in the clustering measurement (the offset between the two could simply be due to the way the data are binned).  Third, it appears that for all data sets the data diverge from the model on scales of a few degrees, such that the fit to the data on smaller scales over-predicts the measurement on large scales.  This discrepancy appears to begin on slightly smaller scales in the unobscured sample. This was not seen in previous measurements (e.g.\ D16).

The biases resulting from fitting the data on scales of $100 < \ell < 1000$ are given in Table~\ref{tbl:results}, and the posterior distributions are shown in the lower panel of Figure~\ref{fig:lensing}.  While the distributions are wider, the two samples clearly have distinct peaks, with a significance of $\sim$2$\sigma$.  The obscured bias distribution overlaps the unobscured by 24\%, and the unobscured overlaps the obscured by 15\%.  

Due to the discrepancy on different scales discussed above, the regions over which we perform the fit to the data impact the final results more so than in previous work.  For example, fitting the same region as D16 ($40 < \ell < 400$), the bias for the obscured sample is reduced to $1.87 \pm 0.18$ and the unobscured bias is reduced to $1.31 \pm 0.15$ --- note that the unobscured sample is affected more strongly.  These measurements are consistent with the others considering the uncertainties, but particularly for the unobscured sample are quite low compared to any previous measurements, including samples selected using other multiwavelength or spectroscopic methods.

Finally, we also note that unlike for the clustering measurements, expanding the footprint of the measurement does not reduce the uncertainty on the final bias measurement, though this does depend on what scales are fit.  While the error bars on individual points are reduced, roughly by a factor of $\sqrt{N}$, where $N$ is the difference in sample sizes, over the $100 < \ell < 1000$ range the final $\chi^2$ values are not significantly changed as the quality of the fit is not improved.  However, no matter which region is fit, the general result that the obscured sample has a higher bias does not change.

\section{Dark matter halo masses} \label{sec:halomass}
Using model fits to cosmological simulations, the large-scale bias of a sample can be converted into the typical or ``effective'' dark matter halo mass of the population.  For a more complete description of this methodology, see D16\footnote{And the code at \url{https://github.com/mdipompe/halomasses}.}. We use the bias parameterization of \citet[][see their Table 2 for the values of the coefficients]{2010ApJ...724..878T}, using overdensity parameter $\Delta = 200$:
\begin{equation}
b(\nu) = 1 - A \frac{\nu^a}{\nu^a + \delta_c^a} + B\nu^b + C\nu^c,
\label{eq:bias_mh}
\end{equation}
where $\nu = \delta_c/\sigma(M)$.  The critical density for collapse of a halo $\delta_c$ is:
\begin{equation}
\delta_c = 0.15(12\pi)^{2/3} \Omega_{m,z}^{0.0055},
\end{equation}
where $\Omega_{m,z}$ is the matter density parameter at $z$ \citep{1997ApJ...490..493N}.  The linear matter variance on the size scale $R_h$ of the halo $\sigma(M)$ is
\begin{equation}
\sigma^2(M) = \frac{1}{2\pi^2} \int P(k,z) \hat{W}^2(k,R_h)k^2 dk.
\end{equation}
The matter power spectrum $P(k,z)$ is calculated with CAMB, and $\hat{W}(k,R_h)$ is the Fourier transform of a top-hat window function of radius $R_h$:
\begin{equation}
\hat{W}(k,R_h)=\frac{3}{(kR_h)^3}[\sin(kR_h)-kR_h\cos(kR_h)].
\end{equation}

The halo masses calculated from the bias measurements are listed in Table~\ref{tbl:results}.  Note that due to the shape of $b(M)$, which flattens at low halo mass and rises steeply at high mass, the uncertainties on $M_h$ are generally asymmetric, and are larger for low halo masses. 

\begin{figure*}
\centering
   \includegraphics[width=16cm]{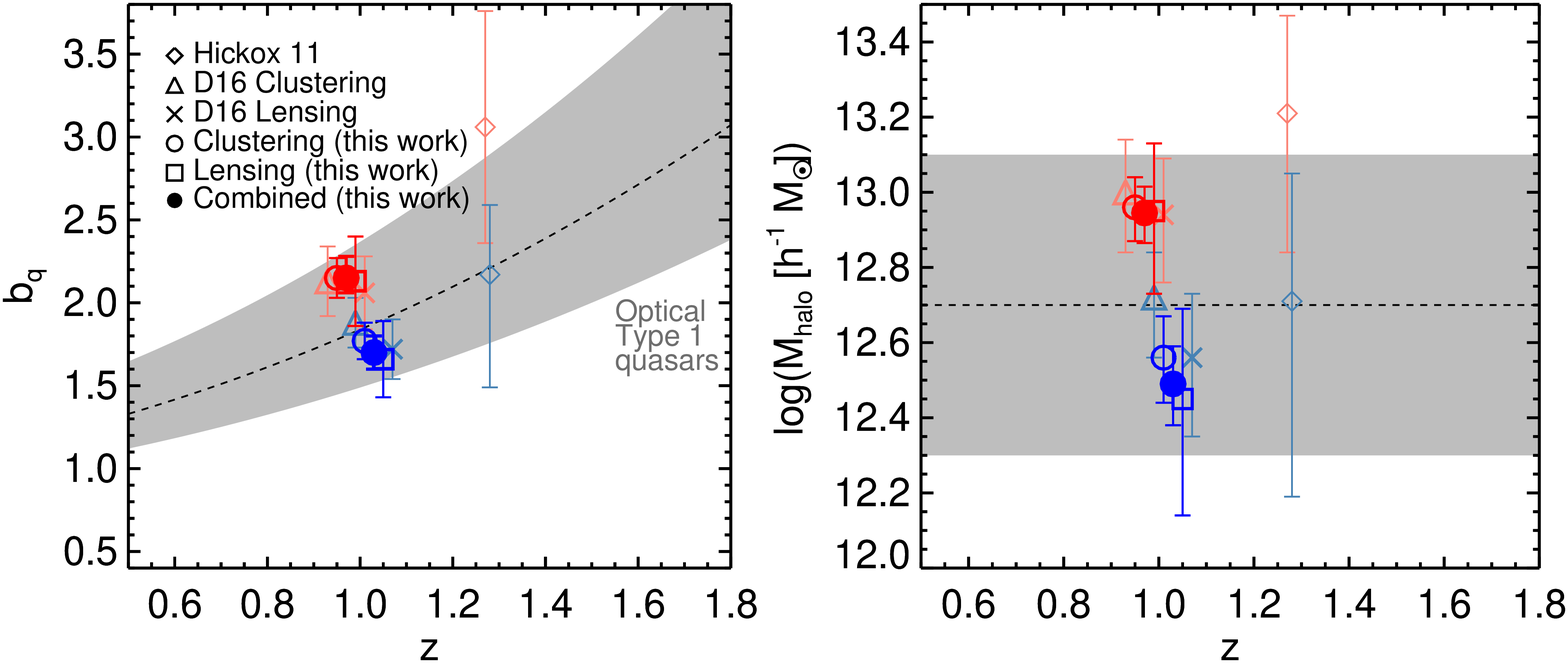}
    \vspace{0cm}
  \caption{The bias (left) and halo mass (right) of quasars across redshift.  Red and blue symbols represent obscured and unobscured quasars, respectively. The grey region shows the typical range for optically detected unobscured quasars from the literature.  The diamonds at $z\sim1.3$ show the results of \citet{2011ApJ...731..117H} from a cross-correlation with galaxies.  The angular clustering (triangles) and CMB lensing cross-correlations (exes) of D16 (light coloured symbols) and this work (circles and squares, darker symbols) are plotted with slight offsets in $z$ for clarity.  The filled circles are the weighted, combined results, representing our final adopted bias values. There is excellent consistency between the results, with obscured quasars always having a higher bias and thus larger halo masses.\label{fig:bias}}
\end{figure*}

\section{DISCUSSION} \label{sec:discussion}
Figure~\ref{fig:bias} shows the results of this analysis (filled symbols) along with other recent work on obscured quasars, including \citet{2011ApJ...731..117H} (triangles), and D16 (light open symbols). Our final bias and halo masses for the obscured and unobscured samples are consistent with previous work (e.g.\ D16) , but the significance of the difference between them is increased. While the measurements all rely on the same underlying data set (the \wise-selected quasars), they are made in independent ways, with different systematics that allow us to combine them for a robust comparison.  Before combining the results, we repeat the CMB lensing cross-correlation analysis in the 100 sub-regions used for the jackknife clustering uncertainties (for the full \wise\ sample), and fit the bias for both measurements in each iteration.  This provides a direct test of the independence of the clustering and CMB lensing cross-correlations.  Interestingly, the final values are not correlated, with $r=0.11$ and $p =0.26$ (using a simple Spearman rank correlation).  This is likely because the error budget in each case is dominated by different sources --- based on the analysis above, it is possible that the clustering result is dominated by Poisson noise, while the CMB lensing is dominated by systematics in either or both the CMB lensing and quasar density maps.  If we combine the measurements here, we find weighted mean values of $b_{\textrm{obsc}} = 2.15 \pm 0.11$ and $b_{\textrm{unob}} = 1.70 \pm 0.10$, or $M_{h\textrm{,obsc}} = 12.94 \pm 0.08$ $M_{\odot}$ $h^{-1}$ and $M_{h\textrm{,unob}} = 12.49 \pm 0.11$ $M_{\odot}$ $h^{-1}$, a difference of $\sim$5$\sigma$. These values are listed in Table~\ref{tbl:results}.

As mentioned in Section~\ref{sec:lensing_results}, it appears that additional systematics in \wise\ quasar selection may become more important over large areas, affecting in particular the CMB lensing measurements.  This manifests in bias uncertainties consistent with or larger than previous measurements and systematically lower bias values, which we mitigate somewhat with the more stringent cut in Galactic latitude.  Interestingly, the additional systematics appear to impact the unobscured sample more strongly than the obscured. 

While we have corrected for the fluctuating \wise-selected quasar density as a function of Galactic latitude, it appears that contamination may become an increasing problem over wider areas.  While \citet{2012ApJ...753...30S} find a reliability of 95\% for their selection criteria in the COSMOS field, it is possible that this is not true across the entire sky.  In addition, it is also the case that simply expanding the area allows more rare contaminants to appear in the sample.  While the amount and type of contamination is difficult to determine from these measurements alone, there are some indications, particularly in Figure~\ref{fig:lens_stack}, that this could be the case.  In general, the trend between $\delta_q$ and $\kappa$ will flatten for smaller samples, as Poisson noise dilutes the signal, particularly in the more extreme bins (this is clearly seen by simply randomly down-sampling one of the samples and comparing it to the full measurement).  However, this is not the case at the low-$\delta_q$ end in Figure~\ref{fig:lens_stack}, where the larger unobscured sample flattens more rapidly than the obscured.  This is in contrast to the high $\delta_q$ side, where they are more consistent.  In a similar figure in \citet{2015MNRAS.446.3492D}, which covered a smaller area, this difference is not apparent.  This suggests that there are more contaminants in the unobscured sample over the expanded footprint, which are scattered randomly across the area so that they are likely to fall in low quasar density pixels, and these are uncorrelated with the CMB.  

The most likely such contamination comes from stars, though we find this unlikely to be the case, at least for typical stars.  As a simple check, we match our unobscured quasar sample with the catalog of \citet{2015MNRAS.452.3124D}, which provides quasar and star probabilities for all point sources in the SDSS, based on their mid-IR through UV fluxes.  Since resolved sources are not included in the catalog, we find 79\% of our unobscured sample in this catalog.  However, comparing the distributions of quasar probabilities $P_{\textrm{QSO}}$ over the full footprint and the region of D16 shows that they are nearly identical, with $\sim$97\% of the sources in both regions having $P_{\textrm{QSO}} > 0.9$. Any stellar contamination is thus likely to be from rare stars with an AGN-like spectral energy distribution (e.g.\ AGB stars), and therefore difficult to disentangle without extensive spectroscopy.  Finally, we also create \textsc{HEALPix} maps at several resolutions of the unobscured to obscured ratio, and find no strongly inconsistent outliers, nor any trend with sky position, suggestive that potential stellar contamination affects both samples equally.

We also explored a more conservative cut of $W1-W2 > 1.0$, for which \citet{2012ApJ...753...30S} report a reliability of nearly 100\%.  This reduces all of the samples by $\sim$40\%.  While such a cut also impacts the redshift distributions of the samples, changing the interpretation of the bias results, we use it simply to check if it improves any of the signs of contamination discussed so far.  The binned $\kappa$-$\delta_q$ relationship does appear to behave more as expected, and the outlier in the unobscured sample in the lowest $\delta_q$ bin is consistent with expectations based just on sample size.  However the CMB cross-correlation bias results remain sensitive to the choice of fitting region, with a seemingly more rapid turnover on large scales, particularly for the unobscured sample.  

Regardless of these issues, we stress that the difference in bias and halo mass for the two populations continues to persist, even if the final adopted values fluctuate.  This is not only bolstered here by the reduced uncertainties on the clustering measurement, but the additional analysis of angular cross-correlations between the \wise-selected samples and with SDSS spectroscopic quasars.  The fact that the \wise\ obscured-unobscured cross-correlation signal lies generally within the individual autocorrelations is strong support for the well-matched redshift distributions based on the Bo\"{o}tes field.  If the samples were mismatched in redshift, we would expect to see a rapid decrease in the cross-correlation signal, as evidenced by the lower amplitude model generated for the cross-correlation with the very similar but not perfectly overlapping Bo\"{o}tes distributions.  Finally, though the uncertainties are large, the cross-correlation with spectroscopic quasars supports the matching in redshift, as well as the final bias results.  All of these taken together are strong evidence that the bias and halo mass difference we find is real.

The most common concerns with these measurements for \wise\ quasars are that the redshift and/or luminosity distributions of the obscured and unobscured populations are not well matched.  The cross-correlations discussed above strongly favor a well-matched redshift distribution, as does the modeling we performed using mock obscured $dN/dz$ distributions at different mean redshifts (Figures~\ref{fig:cross_models} and~\ref{fig:obsc_b_z}).  This, combined with the SED analysis of \citet{2007ApJ...671.1365H} and the analysis of the $W3$ and $W4$ magnitude distributions here (Figure~\ref{fig:w3w4}) strongly suggest a well-matched bolometric luminosity distribution. In addition, as discussed previously, there is little evidence for a strong dependence of the bias on luminosity, implying that even if the samples are at slightly different mean $L_{\textrm{bol}}$, this likely isn't enough to explain the observed difference in halo mass.  However, this may not be the case on small scales \citep[e.g.][]{2017MNRAS.466.3331D}, so care will need to be taken in future analysis of the difference in clustering signal on small scales.

There is general consistency between D16 and this work, especially for the obscured quasars.  The unobscured bias, particularly from the CMB lensing analysis, is somewhat low, both compared to D16 and other work on optically selected unobscured quasars more generally.  Compared to \citet{2011ApJ...731..117H}, we again find general agreement. We also note that our measurement of the overall \wise-selected quasar bias (not shown in Figure~\ref{fig:bias} for clarity, but given in Table~\ref{tbl:results}) agrees well with a cross-correlation with a CMB lensing map from the South Pole Telescope \citep{2013ApJ...776L..41G}.  The interpretation in that work was that the similarity of the overall IR-selected quasar bias to the well-studied optically-selected quasar bias implied that the obscured and unobscured populations were similar.  However, we see here that while the overall \wise\ quasar bias agrees with the mean optical quasar bias, the IR-selected unobscured bias is slightly lower and the obscured bias is slightly higher.

Finally, we point out that this difference in bias and halo mass, supported from many lines of evidence, is likely a lower limit on the value for a truly evolutionarily distinct population \citep{2016MNRAS.460..175D}.  This is simply because the hot dust that makes up the torus is likely to obscure some quasars that are otherwise intrinsically the same as the unobscured population --- i.e. the obscured population is made of of some objects unified by orientation and some unified by evolution or other factors.

Given that extending the survey area seems to lead to an increase in contamination that prevents us from improving the precision of bias and halo mass measurements of quasars selected with simple IR criteria, at least for CMB lensing cross-correlations, even in the era of expanding optical surveys (e.g. Pan-STARRS, LSST) that would allow an obscured/unobscured split over larger areas there may not be much more room for improvement simply be increasing sample sizes.  Of course, such surveys also provide additional depth and time-domain information that will certainly be useful for understanding contamination.  Instead, for now we will look towards smaller samples with redshift information, either spectroscopic or photometric, as well as cross-correlations with increasingly large galaxy and spectroscopic quasar catalogs.

\section{Summary}
Our goal in this work is to improve the precision in the bias and halo mass measurements of \wise-selected quasars by expanding the footprint of previous measurements.  This is to provide the most accurate \wise-selected quasar measurement to date, as well as due to the important but not highly significant difference in halo mass between obscured and unobscured populations identified in previous work.  After correcting for observational systematics that lead to a decrease in the \wise-selected quasar density closer to the Galactic plane, we measure the angular clustering and cross-correlations with CMB lensing maps for both samples.  

The precision of the angular clustering measurement improves as expected. However, the CMB lensing cross-correlations do not improve in precision, possibly due to increased contamination (particularly in the unobscured sample), though it is difficult to say for certain. Combining results from the measurements leads to a difference in halo mass of a factor of 3 at a significance of $\sim$5$\sigma$.  We also perform a cross-correlation of both samples with a sample of spectroscopic quasars, also confirming the halo mass difference.  This implies that the simplest unification models that rely only on orientation are not sufficient to explain the full quasar population.

We also measure the cross-correlation of the obscured and unobscured populations, and find that this gives consistent halo mass measurements.  Using models of the redshift distribution of the obscured population at different mean $z$, we show that the cross-correlations strongly suggest that the obscured and unobscured populations have very similar redshift distributions, both centred around $z \sim 1$.  We also compare the 12 and 22 $\mu$m \wise\ detection fractions and flux distributions and find them to be very well matched, suggesting good agreement in the bolometric luminosities of the obscured and unobscured populations selected in this way.  These analyses highlight that it is fair to directly compare the bias and halo mass measurements of the two samples, even if these parameters evolve with redshift and/or luminosity.

The highly significant difference in obscured and unobscured quasar halo mass reported here, using several lines of evidence along with careful testing of the match in fundamental properties, suggests very strongly that obscuration in at least a subset of IR-selected quasars is tied to the host galaxy.  The simplest explanation for this is a link between the evolutionary state of the galaxy and the availability of material to obscure the central engine.

\section*{Acknowledgements}
We thank Andy Goulding for his valuable input, in particular with regards to the new fitting methods used here.  MAD, RCH, and ADM were partially supported by NASA through ADAP award NNX12AE38G and by the National Science Foundation through grant number 1211096.  MAD and ADM were also supported by NSF grant number 1211112, and MAD, ADM, and SE were supported by NSF grant 1515404.  MAD and RCH were also supported by NSF grant number 1515364 and NASA ADAP award NNX16AN48G.  RCH also acknowledges support from an Alfred P. Sloan Research Fellowship, and a Dartmouth Class of 1962 Faculty Fellowship.

\bibliography{/Users/Mike/Dropbox/full_library}

\label{lastpage}

\end{document}